\documentclass[aip,jcp,amsmath,amssymb,reprint,longbibliography]{revtex4-1}
\pdfoutput=1
\usepackage{graphicx}
\usepackage{cases}
\usepackage[dvipsnames]{xcolor}
\usepackage{upgreek}
\newcommand \rmm[1]  { \textrm{#1} }   
\usepackage{array}
\newcolumntype{L}{>{\centering\arraybackslash}m{3.5cm}}

\begin{document}
\title{What They Don't Want:\\ An Analysis of Brexit's First Round of Indicative Votes}
\author{Thomas Sayer}
\email{tes36@cam.ac.uk}
\affiliation{Fitzwilliam College, University of Cambridge, Cambridge, CB3 0DG, United Kingdom}

\begin{abstract}
Since the result of the 2016 referendum, Brexit has been an unpredictable democratic adventure, the finale of which remains unclear. This year, in the final days of March, parliamentarians seized control of the order paper from the Government and held their own indicative votes in an attempt to break the deadlock. In this paper we analyse the results of this unusual cardinal ballot. We express the various motions in terms of `how much Brexit' they deliver, and employ Monte Carlo in an attempt to determine this from the data. We find solutions which reproduce our intuitive understanding of the debate. Finally, we construct hypothetical ordinal ballots for the various Brexit scenarios, using three different processes. The results suggest that the Government would be more successful taking a softer position, and we quantify this. Additionally, there is some discussion of how tactical voting might manifest itself in the event of such an exercise.
\end{abstract}

\date{\today}
\maketitle

\section{Historical Introduction} \label{sec:intro}
\vspace{-0.2cm}

Some years ago now, the Prime Minister of the day was elected on a manifesto to renegotiate Britain's relationship with Europe and then hold a referendum on the outcome of those negotiations. Although the political class was in favour of maintaining membership, significant parts of the Prime Minister's party took the contrary position. As a result, cabinet collective responsibility was suspended, so those aforementioned individuals who were also members of the Government could publicly campaign against its official position. This description of the 1975 referendum on whether the UK should remain in the European Common Market could, of course, equally apply to the 2016 referendum on the United Kingdom's European Union (EU) membership. Indeed, many of the democratic episodes which mark the United Kingdom's relationship with the EU have a repetitive flavour. In 1993 a Labour amendment to the Treaty of Maastrict (Social Protocol) resulted in a tie,\footnote{Although it was later found not to have been} which required Speaker Boothroyd to cast her deciding vote with the Noes.\cite{Hansard} In 2019, Speaker Bercow likewise cast his vote with the Noes to defeat a motion which sought to hold a 3$^\rmm{rd}$ round of \textit{indicative votes} on Brexit. In this paper, we will examine the results of the first round of that rather unusual process, in a period which has gone rather further than merely repeating history.

The 1975 referendum went the `right' way for Harold Wilson. In the 2016 recapitulation, David Cameron was not so fortunate: the following day he resigned with immediate effect. The resignation of the Prime Minister triggered a leadership contest within the Conservative Party. The favourite, Boris Johnson -- who had backed the campaign to leave the EU after a considerable amount of drama\cite{Watt} -- withdrew before the first round following a political bushwhacking from his close ally Michael Gove. The purportedly\cite{Dominiczak} Remain-supporting member for Maidenhead cruised to victory. 

For quite some time afterwards, much of the commentary indulged itself in speculation as to what `kind of Brexit' the Government would negotiate.\cite{AlexandraSima,Kama,J.P.} A hard Brexit, or a soft Brexit? For a long time there was nothing more substantive from Downing Street than tautological statements in the vein of `Brexit means Brexit'. It was not until the 12$^{th}$ of June 2018 that the white paper known as the `Chequers Plan' was released. Within a month, the Brexit Secretary David Davis and the Foreign Secretary Boris Johnson had resigned in protest. In the meantime, the Prime Minister (after explicitly stating she had no intention of doing so) called a snap election which resulted in the loss of the Conservative majority. At the time of writing, the Conservatives remain a minority government, supported by a confidence-and-supply arrangement with the Eurosceptic Democractic Unionist Party (DUP) of Northern Ireland. In spite of this, the negotiators kept calm and carried on. At the Cabinet of the 14$^{th}$ of November 2018 the `Draft Withdrawal Agreement' was debated. The Brexit Secretary Dominic Raab and Work and Pensions Secretary Esther McVey resigned over it. Again, in spite of this,``May's Deal'' was jointly agreed with the EU and tabled to be ratified by parliament -- something that the Government had argued was unnecessary, but the Supreme Court found otherwise.\cite{U}

\begin{figure*}[t]
\begin{center}
\resizebox{0.85\textwidth}{!}{\includegraphics[trim=2cm 1cm 0.5cm 1.5cm]{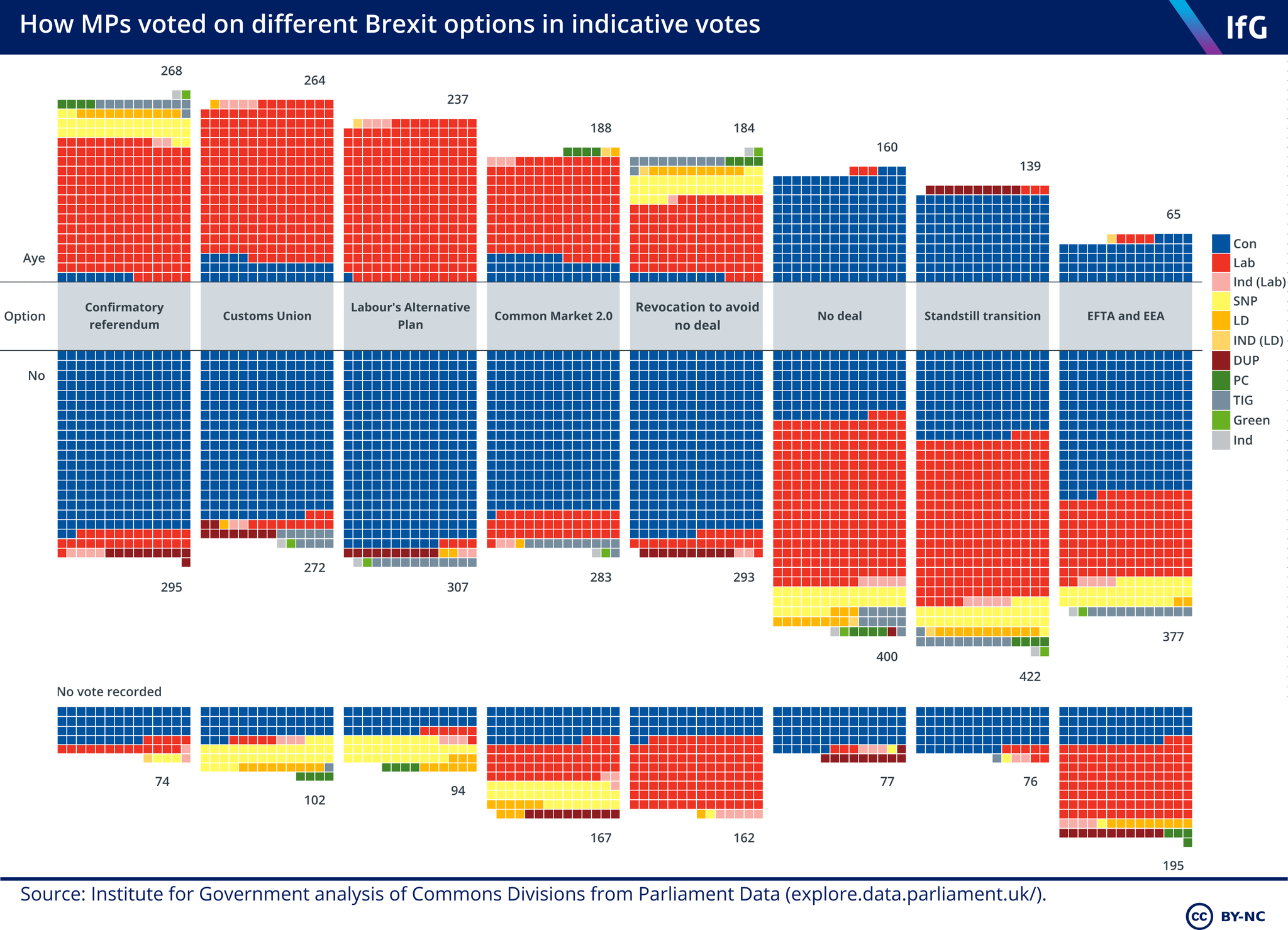}}
\end{center}
\caption{Graphic displaying the party political makeup (see the colour key) for each of the indicative votes of the ayes, noes, and those who abstained. As the reader will no doubt be aware, no motion gained a majority. N.B. `Standstill transition' is ``Planned No Deal'' in the language of Table~\ref{tb:cand}. Reproduced from Ref.~\citenum{Thi} under the CC BY-NC licence.\footnote{https://creativecommons.org/licenses/by-nc/3.0/legalcode}} \label{fig:ifg}
\end{figure*} 

\begin{table}[h] 
\begin{tabular}{|c|L|c|c| } 
\hline
Letter & Official Name & Nickname & Colour \\\hline\hline
B & No Deal & No Deal & blue \\\hline
O & Contingent Preferential Arrangements & Planned No Deal & orange \\\hline
H & EFTA and EEA & Norway & green \\\hline
Z & / & May's Deal & red \\ \hline
D & Common Market 2.0 & Norway+ & purple \\\hline
J & Customs Union & Customs Union & brown \\ \hline
K & Labour's Alternative Plan & Customs Union+ & pink \\ \hline
M & Confirmatory Public Vote & Public Vote & silver \\\hline
L & Revoke To Avoid No Deal & Revoke Article 50 & gold \\\hline 
\end{tabular}
\caption{A list of the eight amendments, with their corresponding letter, the nickname with which they shall be referred to, and their corresponding colour-coding in the various figures that follow. The Government's Withdrawal Agreement is also included for completeness, and given a letter-reference of `Z'.} \label{tb:cand}
\vspace{-0.5cm}
\end{table}

The last major event antecedent to the ratification vote was an humble address. The Government was required to publish its legal advice on ``May's Deal''. Although the motion passed, the documents were not immediately forthcoming. As a result, a vote to hold the Government in contempt of Parliament carried by 311 to 293.\cite{Leilanathoo} Constitutionally unprecedented, they had now gone further than repeating history. In the wake of this defeat, the vote on ratification was surreptitiously delayed. This prompted votes of no confidence: first within the Conservative Party, and then within the House of Commons itself. Both failed.

``May's Deal'' would be voted on for the first time on the 15$^\rmm{th}$ January 2019. It went down by 202 ayes to 432 noes, the largest defeat of a government motion. Ever. A month later, on St. Valentine's Day, the Government was again defeated on a motion in support of their negotiating strategy. Another month later, on the 12$^\rmm{th}$ of March, ``May's Deal'' returned to the Order Paper, reinforced by ``meaningful legal assurances'' from Brussels. It secured 40 more ayes, but still went down as one of the most brutal defeats on record. Unbowed, the Prime Minister secured a technical extension to the Brexit deadline, which at this point was the 29$^\rmm{th}$ of March, in order to table a 3$^\rmm{rd}$ vote. It was at this point that a cross-bench group of MPs under Sir Oliver Letwin amended a Written Statement to temporarily suspend standing orders such that control of the Order Paper was wrested from the Government.

Under Letwin \textit{et al.}, a set of of non-binding motions were collectively debated. Each motion was put to a deferred division and voted on in the normal way. They are listed in Table~\ref{tb:cand}. For clarity, the motions will be referred to by their nicknames enclosed in quotation marks. As I am sure the reader is aware, no motion received a majority. While a second round of voting was carried out, it too failed to reach a consensus; it will not be analysed in this paper. It stands to reason that given that MPs and pundits were confident the first round of voting would not return a clear consensus, the data should be a good indicator of preference. Surprisingly, I could find only one analysis\cite{Thi} (see Fig.~\ref{fig:ifg}) of the unprecedented dataset generated by the indicative votes process.

\section{Representing These Data} \label{sec:rep}
The first thing to do is come up with a good representation of the data. The most obvious way of doing this is the for/against split (by party) of every deal as seen in Fig.~\ref{fig:ifg}. We would like to go further, and be able to see information about each individual MP. Each has their own voting record, and these could be quite interesting. It is known that some MPs chose to \textit{plumper vote}: vote for a single option. Some MPs voted no to everything. Some voted in favour of many things. Furthermore, we know that some amendments (henceforth referred to as \textit{candidate policies} or \textit{candidates} for short) were considered more similar in terms of their outcome than others. For example, ``Customs Union+'' and ``Public Vote'' are both Labour Party policy; it would not make a lot of sense if policies seen as consistent (at least by Labour) were viewed as leading to vastly different outcomes. There is therefore some sense of a `degree of Brexit': the extent to which a candidate advocates moving away from the EU. We shall refer to this as \textit{hardness}. To reduce the debate to this one measure would be a \textit{single-axis  projection}. A familiar single-axis would be the common parlance of left- and right-wing. To say that someone is `right-wing' (aside from revealing your political definition of the center) is to reduce all their politics to a single value. Moreover, this does not always represent the same axis over time, even for a fixed set of possibilities in the larger space (where all one's policy positions are judged on their own axis). 

\begin{figure}[h]
\begin{center}
\resizebox{0.4\textwidth}{!}{\includegraphics[trim=0cm 0.5cm 0cm 0.5cm]{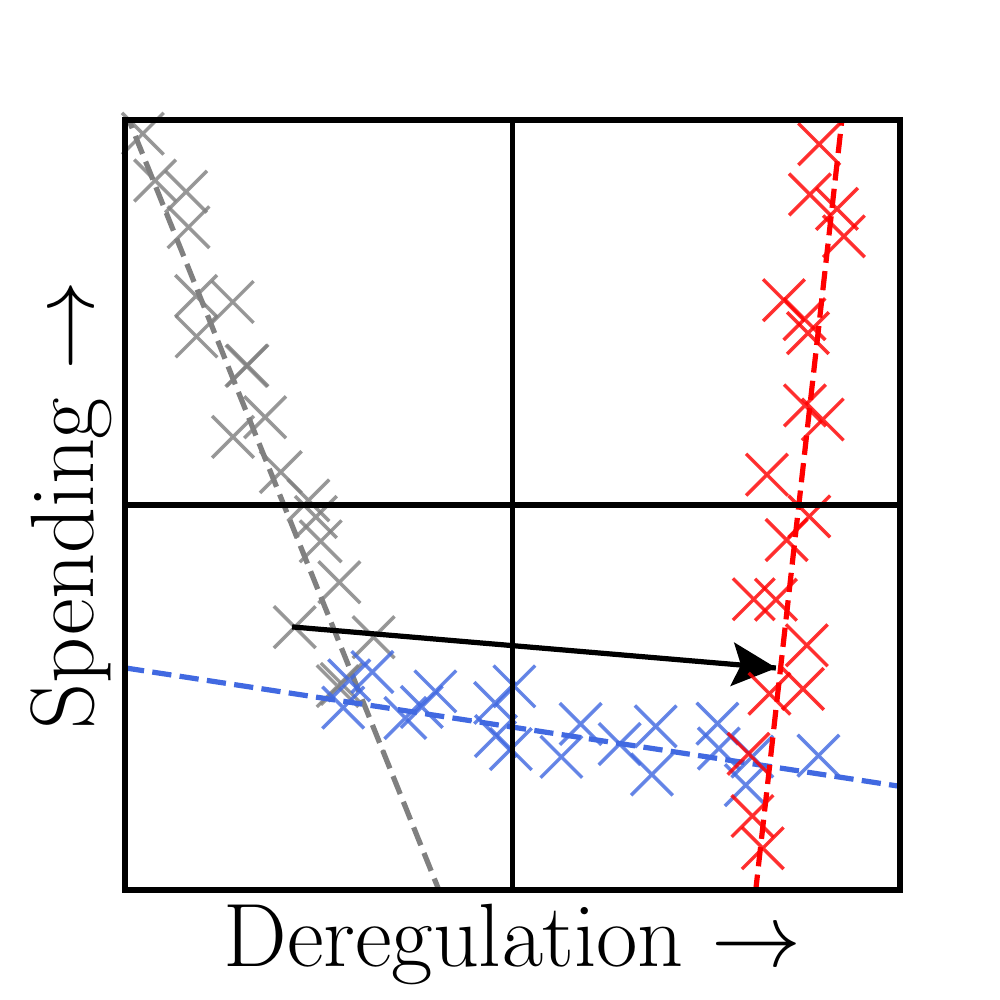}}
\end{center}
\caption{Schematic representation of how multidimensional data are represented by a single axis. The arrow shows how the projection of a point in the grey set (postwar period) translates onto the red axis (New Labour). The blue represents the Thatcher period. The figure is intended to be illustrative. The scatter of the points and the parameters of the lines are not meant to be quantitative representations of the historical details, mostly because I have forgotten them.} \label{fig:pax}
\end{figure}

For example, in postwar Britain there was an economic consensus that the state would play a central role in the economy, and the debate was as to the extent to which Keynesian stimulus should be applied. Roughly speaking. With the rise of Thatcherism, there was a shift in what was considered to be acceptable (often referred to as the \textit{Overton window}). The role of the state was drastically reduced, and so was spending. Someone who was considered right-wing during the postwar period would, having not changed their position, be considered left-wing under Thatcher. In the neoliberal hellscape we now exist in, private markets are taken as given, while the left-right debate focuses once again on the degree of public spending.\cite{King2013} That same individual, having not moved in 50 years is, upon being projected all the way onto the neoliberal axis, once again right-wing. This is shown schematically in Fig.~\ref{fig:pax}.

To perform this single-axis projection, we first have to decide on a numerical hardness value for each candidate. To begin with we shall make the simplest assignment: labelling the entries in Table~\ref{tb:cand} in integer steps from -4 to 4. This spacing is not going to be optimal. Furthermore, the ordering is (at this point) entirely an opinion based on my subjective understanding of the debate. In Sec.~\ref{sec:cardinal} I shall try to address these shortcomings. In any case, given a \textit{hardness set} for the candidates we can calculate each MP's mean hardness value, by averaging over their voting record. We shall count votes aye as 1, abstentions as $\frac{1}{2}$, and votes no as 0. This will categorise each MP on the Brexit axis: a score of -4 means an extremist \textit{Brexiteer}, while a score of +4 indicates an extremist \textit{Remaineur}. For intermediary values, however, several different voting patterns can return the same average hardness. Casting votes symmetrically about a candidate will look the same as voting only for that candidate. We want to be able to visualise this difference, so we will also assign each MP a score based on the variance of their voting record: Someone with a spread of 0 is a true \textit{ideologue}, willing to tolerate only a single candidate; someone with a high spread is a \textit{pragmatist}. We will discuss tactical voting later, but in this section we will subsume it into our definitions: the plumper votes for ``No Deal'' are not seen in a tactical light, but rather as the true expression of those MPs' distaste for any other option. In the Appendix, Fig.~\ref{fig:original} shows these voting data before any manipulation.
\\

\begin{figure*}[t]
\begin{center}
\resizebox{0.75\textwidth}{!}{\includegraphics[trim=1cm 1.5cm 1cm 1.5cm]{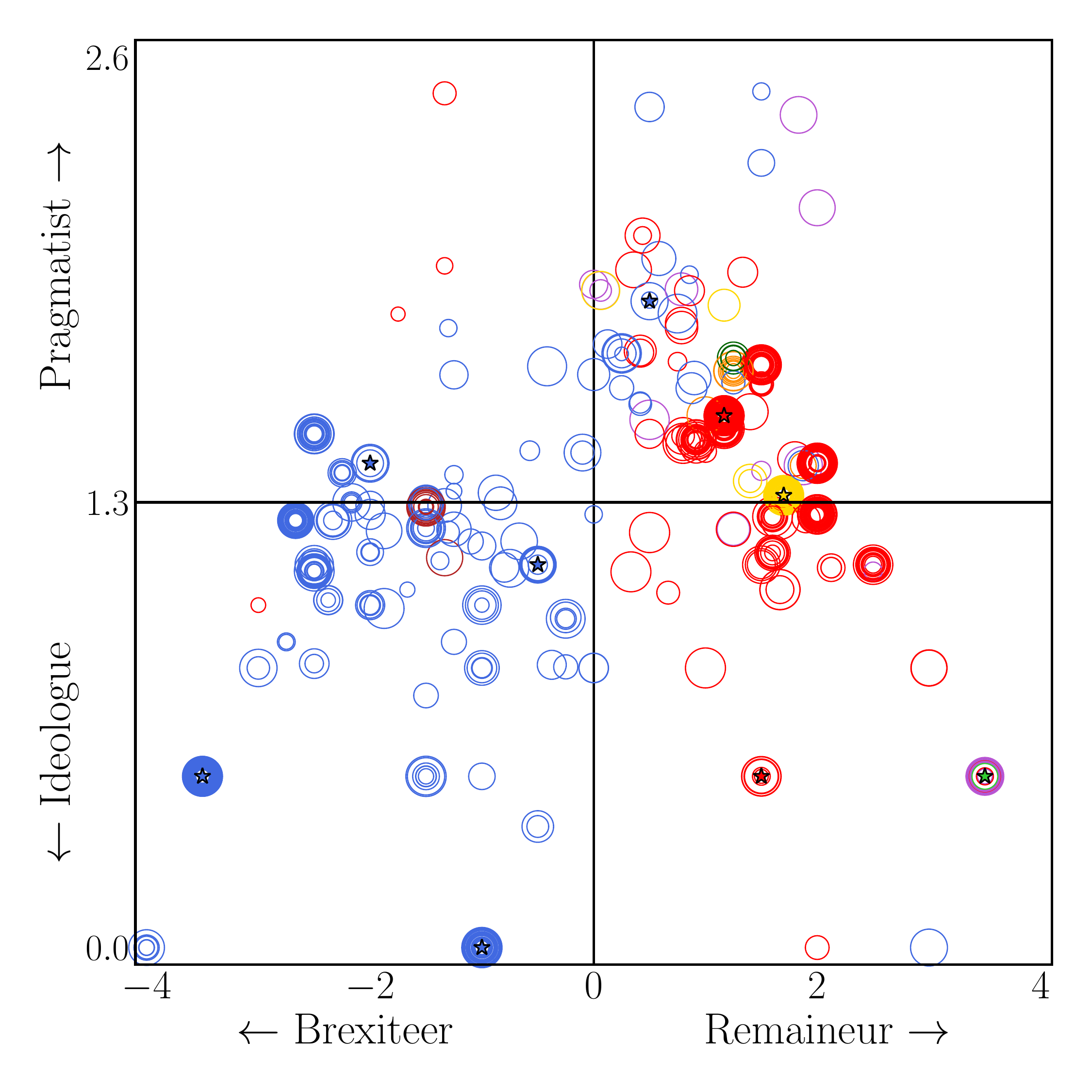}}
\end{center}
\caption{MPs are represented by circles of random radius and coloured according to their party at the time, as in Fig.~\ref{fig:ifg}. N.B. all independents are in purple, whether or not they aligned with The Independent Group (TIG). Points in the blank areas on the left and right of the main distribution are (generally) not representable by any possible ballot paper. Labour MP Stephen Hepburn is not visible as he voted no to all options and to ``May's Deal'' (on all three occassions). Huw Merriman voted for ``No Deal'' and ``Public Vote'', as such he has a variance of 2.87 and has been excluded for clarity. The stars represent (from -4 to +4): Marcus Fysh, George Eustice, Chris Skidmore, Oliver Letwin, Antoinette Sandbach, Jeremy Corbyn, Rosie Cooper, Joanna Cherry, and Caroline Lucas.} \label{fig:bigplot}
\end{figure*} 

There are two major problems to be immediately addressed. Firstly, the Cabinet, Sinn F\'{e}in, the Speaker and his Deputies, and 1 SNP MP, are considered to have abstained on all options, appearing centrally with (ironically) a high degree of pragmatism. For now we will simply remove these MPs from the set. Secondly,  there are a large number of MPs who voted no to every option: 32 of them Conservatives, 1 Independent, and 1 Labour. We currently know nothing about them. There are two reasons I can think of for their doing this:
\begin{itemize}
\item They hold the whole issue of Brexit in contempt and refuse to participate.
\item They are ``May's Deal'' ideologues, but were unable to express their view due to the form of the ballot paper.
\end{itemize}

That the lack of ``May's Deal'' from the ballot paper is causing more people to appear as pathologically ideological than in reality exist is concerning for our interpretation. To resolve this, I propose that we include data from the votes on the United Kingdom's Withdrawal from the European Union (i.e. ``May's Deal''). This data is obviously somewhat tarnished by the imposition of whipping. However, the whip was somewhat ineffectual, at least on the Conservative side. I have therefore included the average of the first two votes (which were very similar, but help separate the different MPs currently sitting at (-4,0) into different groups, see Fig.~\ref{fig:mayvotes}) for each MP.

\begin{figure*}[t]
\begin{center}
\resizebox{0.7\textwidth}{!}{\includegraphics[trim=3cm 2cm 3cm 0.2cm]{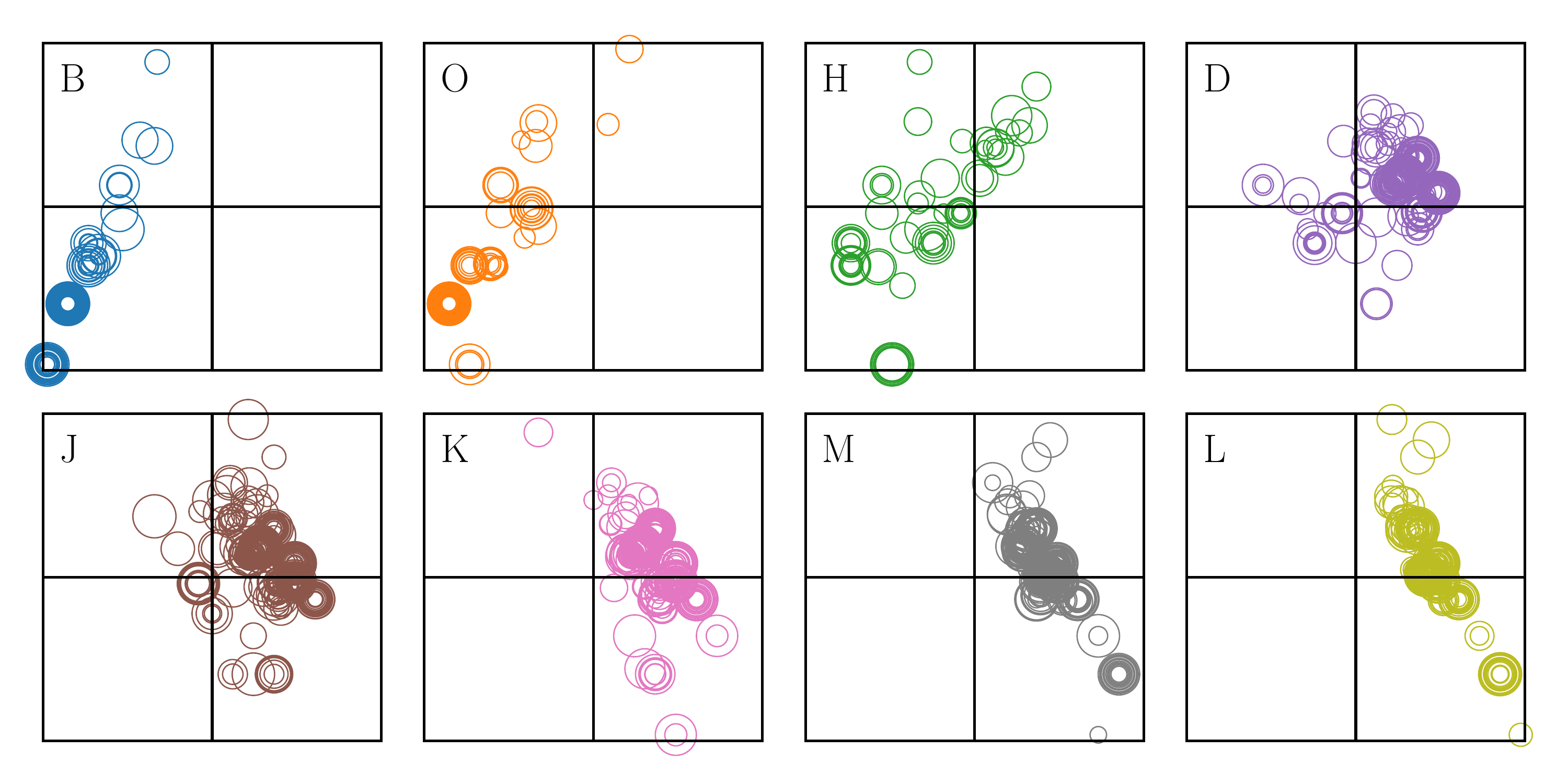}}
\end{center}
\caption{Fig.~\ref{fig:bigplot} where only those who voted aye to the indicated motion are plotted. This is complimentary to Fig.~\ref{fig:ifg}. It stands to reason that motions which display similar patterns should be clustered together in the optimised hardness set.} \label{fig:allyeses}
\end{figure*}

Having done this, the final values are displayed in Fig.~\ref{fig:bigplot}, colour-coded by political party. In the Appendix, a breakdown of each candidate's supporters is displayed in Fig.~\ref{fig:allyeses}. Here we can see that the 4 most extreme candidates had quite narrow voting bases. ``Norway'' managed to draw the broadest support from within the Conservative Party, but failed to crack the ideological far-left, which is ironic since those are the people who voted Leave in the referendum (as can be seen in Fig.~\ref{fig:refer}) and it was exactly those people who originally advocated for the Norway model (without quotes). ``Norway+'' and ``Customs Union'' were the only candidates to draw truly cross-bench support, as could be seen in Fig.~\ref{fig:ifg}. Here, the fine-detail of exactly what kind of people voted for one and not the other can be readily observed in the 2D representation. Moreover, the (dis)similarity between the support bases of various motions motivates us to find a better hardness set. A final comment is that the DUP appear as only mildly left of centre, despite never voting for ``May's Deal'', even in the third vote. This is likely because they place considerably more value on how each candidate will deal with the border on Ireland, and this single-axis projection loses the information.

\vspace{-0.2cm}
\section{Inference from Cardinal Preferences}  \label{sec:cardinal}
\vspace{-0.1cm}

The indicative votes process is different from other votes in parliament. When we elect MPs in Britain, the candidate who gain the most votes is duly elected; this is known as \textit{plurality voting}. In parliament itself, the principle is the same except there are only two options: votes to abstain are the same as not having been entitled to vote at all (or indeed as voting in both lobbies). However, once the parliamentary `domain' of \{aye, abstain, no\} is extended to consider multiple solutions at the same time, it becomes possible to express more information about your preferences by using the abstention. This is known as \textit{cardinal voting}. The common government line around this period was that parliament had to stop saying what it didn't want (voting down ``May's Deal'') and start saying what it did want. This was a salient observation. This goes one step further in that it was also possible to say what one didn't really mind. (Although when this method is actually used to determine a winner, game-theoretic analysis demonstrates that casting abstaining votes is a dominated strategy and should not be pursued, although of course, people are still observed to pursue it.\cite{Felsenthal1989}) Indeed, at the time of the first round of votes, Liberal Democrat MP Layla Moran appeared on the BBC's \textit{Politics Live} explaining that she interpreted the indicative vote ballot paper as asking for her opinion on each candidate on a scale of $\lbrace1,0,-1\rbrace$. As it happens, by considering abstentions in the way I have done, I have also adopted this interpretation and merely shifted and rescaled it to $\{1,\frac{1}{2},0\}$. 

\vspace{-0.4cm}
\subsection{Single-Peaked Preference Model}
\vspace{-0.2cm}

\begin{figure}[h]
\begin{center}
\resizebox{0.4\textwidth}{!}{\includegraphics[trim=0cm 0.65cm 0cm 0.75cm]{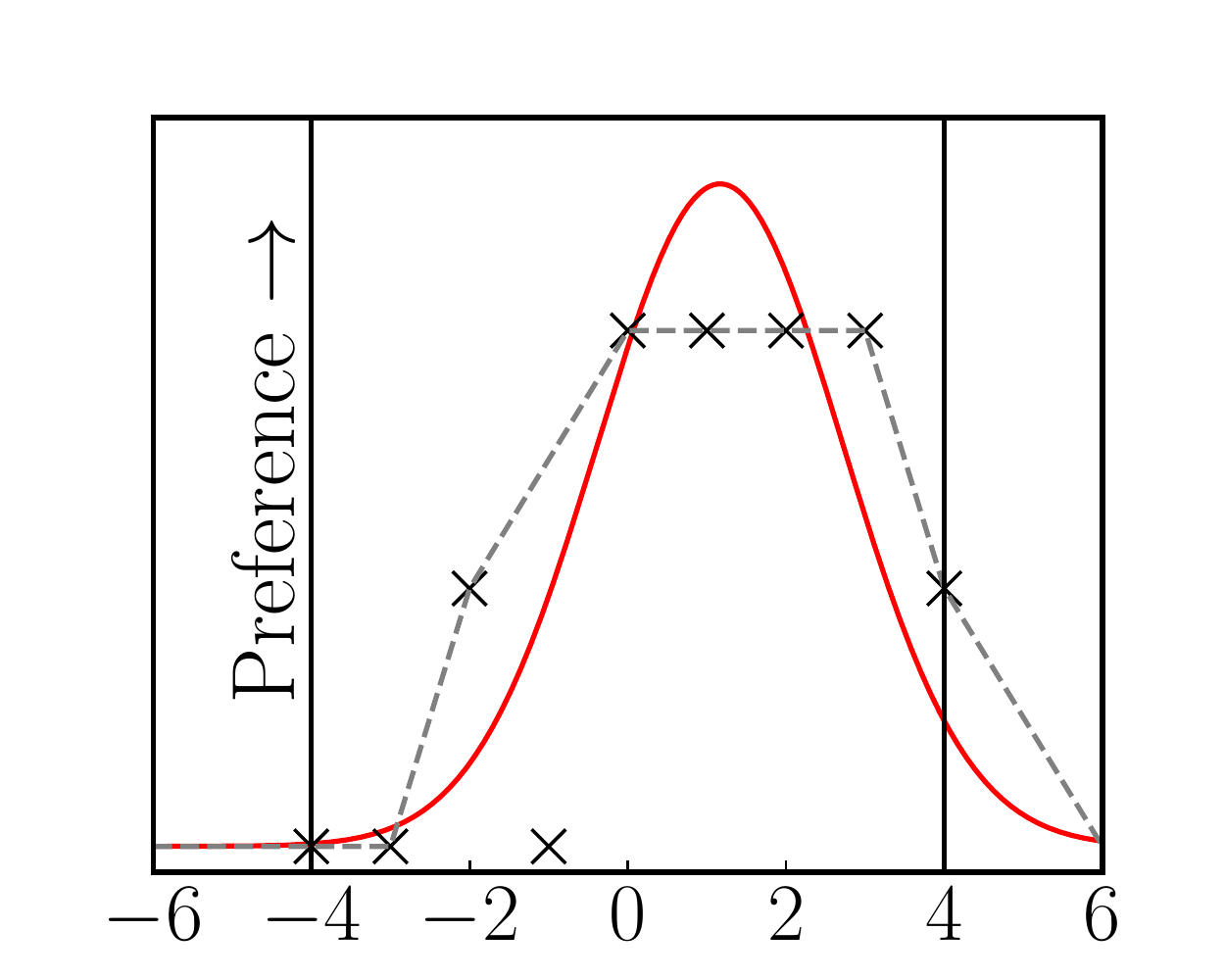}}
\end{center}
\caption{Example of a logical MP (red curve). The black Xs represent the voting record (0, $\frac{1}{2}$, or 1) to each of the proposals spaced along the x-axis. The dotted grey line is to guide the eye. In this case the curve well-represents the underlying data. The inclusion of the vote on ``May's Deal'' at -1 had no affect on this fit, as it was a no vote.} \label{fig:JC}
\end{figure}

We now ask what can be gleaned from these cardinal preferences by way of modelling. Specifically we are searching for some clue as how to treat the definition of the Brexit axis in a more quantitative and objective manner. At this point we have obtained, for each MP who did not abstain on all options, a mean and a variance. We now posit that each MP has a \textit{position} on Brexit: a favoured hardness of Brexit that is described by the mean of their voting record. For any other candidate, the desirability of that policy is less the farther away it is from the mean. We describe this behaviour \textit{via} a Gaussian whose variance is exactly the variance of their voting record. We refer to these curves as \textit{logical MPs}, a phrase that should allow for no ambiguity between the real politicians and the mathematical functions. An example is given in Fig.~\ref{fig:JC}. 
One observation about the resulting profile is that it can return non-zero values for positions outside of the Overton window. This can be interpreted without too much difficulty. Positions to the right of ``Revoke Article 50'' correspond to adopting a more integrationist policy towards the EU than we had previously followed; for example, 5 might correspond to adopting the Euro. Positions to the left are a little more fanciful given how far ``No Deal'' already is from our current position, but it is still possible to conceive of policy positions which advocate adversarial attitudes towards the EU; -5 might be Foreign Secretary Mark Francois starting a trade war with Spain over Gibraltar, or something equally exciting. A further observation is that logical MPs have preference curves that are symmetrical about the mean value. This property is not going to be satisfied by most of the MPs in the original set, but owing to the small size of the data available it would be taking liberties to calculate any higher moments. Later in this section, when we address the definition of the Brexit axis, we will go some way to accounting for this. 

This rather natural way of expressing preference is quite appealing, and so it makes sense that someone else thought of it before me. Indeed, a cursory Google search returns Marie Jean-Antoine-Nicolas de Caritat, Marquis of Condorcet. My understanding is based on Hillinger's\cite{hillinger} reporting of Condorcet's model, as the original 1785 work\cite{Condorcet1785} is in French, and my attempt at translating it was somewhat unreliable. Condorcet seems to be to Thomas More as Mathematics is to Theology. Unfortunately, he was so clued-up that Robespierre quickly realised he had to murder him to death. The Republic had no need of scientists. Nevertheless, Condorcet gives his name to a family of voting methods which seek to solve a problem like this one by obtaining the candidate who would defeat all others in pairwise contest. We will return to this in the next section.

For now, we shall ask our logical MPs to take part in what is known as a \textit{cumulative vote}. We normalize each curve to 1, that is to say, we give each logical MP a single vote which they may split up and distribute across the axis in accordance with their preference for that particular hardness (our logical MPs will vote \textit{sincerely}). Our extremists assign their whole vote to a single point on the axis (their curve is a delta function), while our pragmatists split their vote across many options. We then sum over all MPs to obtain a distribution that represents, in some sense, the support for a Brexit of each particular hardness. We term this the \textit{popularity profile}. The integral under the curve is equal to the total number of MPs involved.

However, before we perform this exercise, we wish to take into account the Cabinet. The Cabinet are large enough that we cannot discount them and still draw conclusions about the relative strength of support between candidates. Hopefully we can account for their hypothetical behaviour by suggesting that they are, to some extent, representative of the Conservative Party as a whole. One obvious way to implement this would be simply to assign each and every member the average mean and variance of all the Conservative MPs. This would be rather suboptimal, especially given that the range of views known to exist in the cabinet is quite broad. At the other end of the spectrum, Philip Hammond has suggested that a second referendum would be `perfectly credible'. On the other end, after a hard week spent not scheduling opposition days and dodging the buckets collecting water in the dilapidated Palace of Westminster, Andrea Leadsom presumably arrives back at her constituency home, puts on a brew, swaps into her ruby slippers, and gets stuck-in to the latest ERG memorandum.  

\begin{figure}[h]
\begin{center}
\resizebox{0.4\textwidth}{!}{\includegraphics[trim=0cm 1.0cm 0cm 0cm]{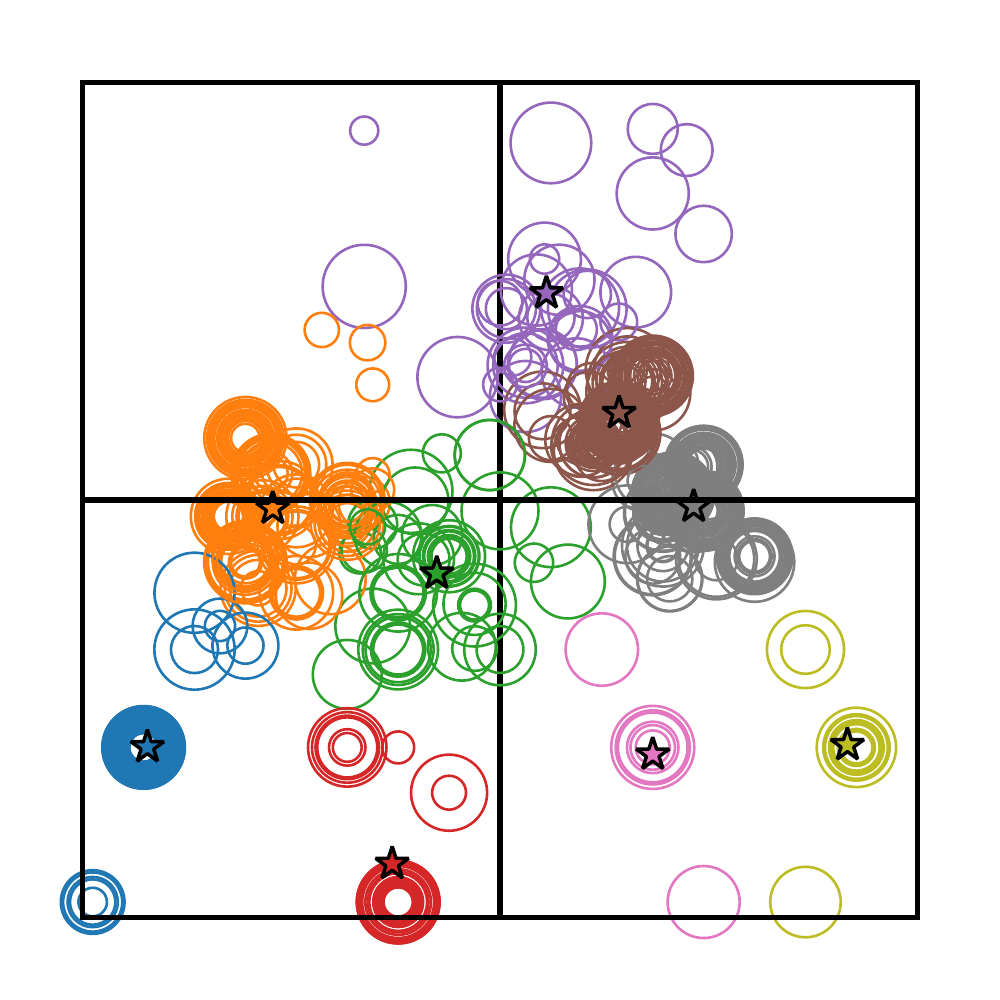}}
\end{center}
\caption{Voronoi partition (on the displayed, rather than numerical, norm) of Fig.~\ref{fig:bigplot} where the centers used in the construction were the MPs highlighted in that figure. Those MPs are listed and colour coded in Table~\ref{table:voronoi}, below. In this figure, the stars represent the average mean and variance of the MPs within the corresponding cell (coloured grouping).} \label{fig:voronoi}
\end{figure}

In order to reflect these differences we will identify characteristic groups of MPs.\footnote{Alternative title: `Machine Learning and the Maybot'} Students of secondary school history may recall one John Snow who, knowing quite a lot, noted during the 1854 Broad Street cholera outbreak that the vast majority of the victims lived closer to that street's water pump than to any other. By demonstrating infected water was the source, he thus disproved the miasmatic theory of contagion, at least for cholera. This sort of distance-based categorization is known as a Voronoi partition. Here, we perform the same exercise on these mean and variance data with respect to the singled-out individuals in Fig.~\ref{fig:bigplot}. That is to say: the logical MPs are the citizens and the starred individuals are the water pumps. The result of this is displayed in Fig.~\ref{fig:voronoi} by a colour-coding, where the new set of stars represent the average mean and variance within each group. The members of the Cabinet may then be distributed amongst these groups, proportional to what fraction of the total parliamentary Conservative Party exists in each. The results are displayed in Table~\ref{table:voronoi}.

\begin{table}[h]
\centering
\begin{tabular}{ |c|c|c|c| }
\hline
Voronoi Center & Assignment & Colour & Share\\
\hline\hline
Marcus Fysh & (-3.46, 0.50) & blue & 8 \\\hline
George Eustice & (-2.23, 1.27) & orange & 13 \\\hline
Chris Skidmore & (-1.06, 0.12) & red &  4 \\\hline
Oliver Letwin & (-0.62, 1.06) & green & 5 \\\hline
Antoinette Sandbach & (0.46, 1.97) & purple & 2 \\\hline
Jeremy Corbyn & (1.17, 1.58) & brown & 1 \\\hline
Rosie Cooper & (1.50, 0.48) & pink & 0 \\\hline
Joanna Cherry & (1.90, 1.28) & silver & 0 \\\hline
Caroline Lucas & (3.41, 0.51) & gold & 0 \\
\hline
\end{tabular}
\caption{A list of the MPs used in the Voronoi partition, the average mean and variance of their resulting grouping (the stars in Fig.~\ref{fig:voronoi}), the colour assignment, and the number of Cabinet members expected to be in that grouping by proportionality.} \label{table:voronoi}
\end{table}

\begin{figure*}[t]
\begin{center}
\resizebox{0.8\textwidth}{!}{\includegraphics[trim=0cm 1.4cm 0cm 2cm]{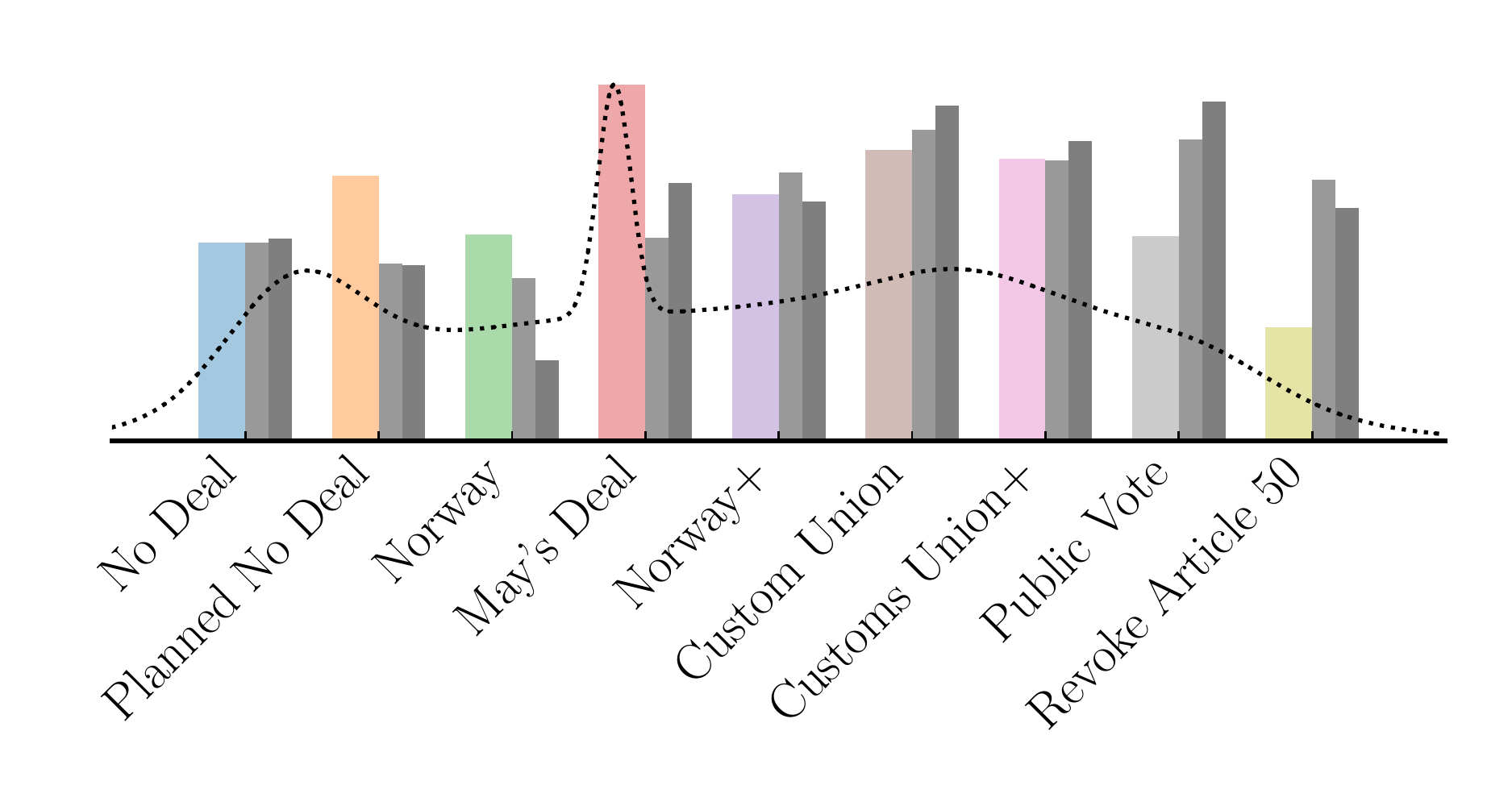}}
\end{center}
\caption{The coloured bars represent the model data of Fig.~\ref{fig:op0fine} integrated according to the rule described in the text. The adjacent light-grey bars are the original data, with abstentions scoring at 0.5, and with the sum over all the candidates normalized to the same total as the coloured bars. The final set of dark grey bars do not include the abstentions, this is for reference. The major difference is on the popularity of ``Norway''; conclusions about the most popular options are largely unaffected.} \label{fig:op0}
\end{figure*}

\begin{figure}[h]
\begin{center}
\resizebox{0.5\textwidth}{!}{\includegraphics[trim=0cm 1cm 0cm 0cm]{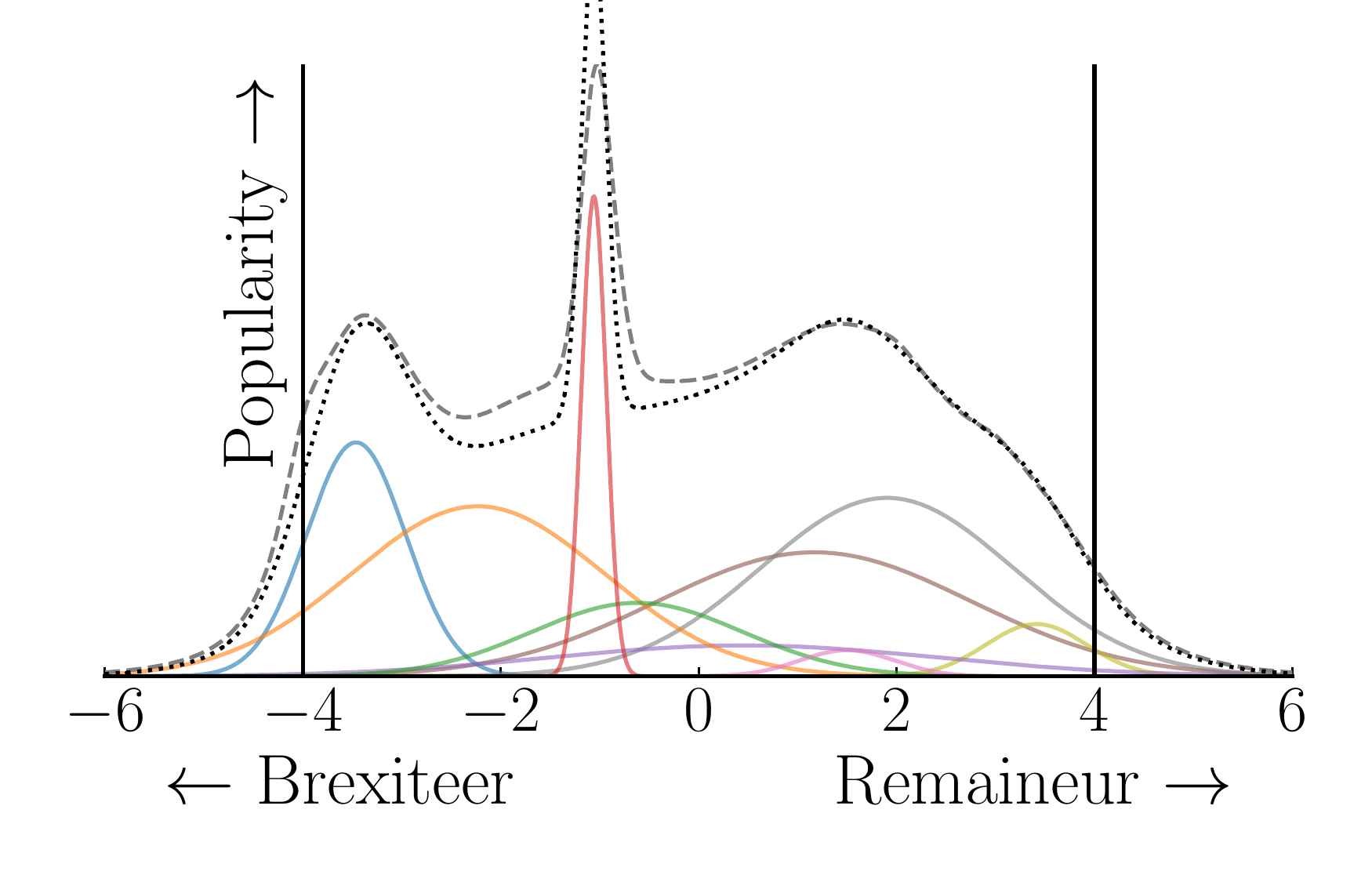}}
\end{center}
\caption{Continuous popularity profile obtained from summing over the logical MPs (grey dashed line). The groups of Fig.~\ref{fig:voronoi} are also displayed, and their sum is the black dotted line. The delta-like spike around -1 is due to the ``May's Deal'' ideologues, who are assigned a minimal spread (such that their logical MP goes to zero as it reaches the adjacent options) for clarity. The group-average variance is even smaller than this.}  \label{fig:op0fine}
\end{figure}

We are now in a position to construct the popularity profile of Fig.~\ref{fig:op0fine}. The distribution has three principle peaks corresponding to: ``No Deal'', ``May's Deal'', and the ``Customs Union/+'' options. This was already evident by inspection. If we were to pick the most popular point it would be the peak corresponding to ``May's Deal''. Sadly, the real candidate-space is not continuous like this, or, in other words, the width of the peaks are also important. To project onto the discrete (original) space we integrate over the Brexit axis for each candidate such that the limits of integration are defined as half-way between that candidate and its left/right nearest neighbour. This means that the extreme options, ``No Deal'' and ``Revoke Article 50'', capture all support to the left/right of them, respectively. Having performed this projection, we can now compare our model values with the real-life results of the indicative vote process. These are the coloured and light-grey bars in Fig.~\ref{fig:op0}. The dark-grey bars are the aye-votes only, for reference. As the figure shows, support for some options like ``No Deal'' and ``Customs Union+'' are almost perfectly reproduced by this model. However, some have a large degree of error, such as ``May's Deal'', ``Public Vote'', and ``Revoke Article 50''. We can quantify how well the model has performed by calculating the modulus of the vector containing each of these errors. For this example the value is 0.124 in relative units: around 12\% of the votes are mis-distributed. 

The reason for performing this exotic and hypothetical form of voting will now be explained. In the first section, we bemoaned the fact that the Brexit axis' spacing was arbitrary, and its ordering subjective. To some extent, the spacing and ordering of the axis also subsumes the issues we discussed around the Gaussian representation of the logical MPs: while they cannot have any asymmetry, the position of the options around the peaks can. It is now claimed that the definition of each candidate's hardness is the major source of the 12\% error in the model. As far as possible, the model should be self-consistent, since it takes parameters from the data set and then attempts to reconstruct that same data set. One issue that I will not be able to resolve is that of MPs who view the candidates in a completely different order from the majority. For example, Conservative Huw Merriman voted aye to ``No Deal'' and ``Public Vote'', apparently because he has more faith in the democracy than ``No Deal'' supporters.\cite{HuwMerriman} This will lead to a residual level of error we cannot overcome. Nevertheless, a different choice of hardness values for the candidates (hardness sets) should have a lower error, and if we could find such a set we could go some way to commenting on the `objective' spacing and ordering of options. Note, it is not so simple as to simply move, for instance, the ``Revoke Article 50'' candidate further into the large peak of support just adjacent to it! Altering the hardness of a particular candidate affects the means and the variances of all the MPs who voted for it, and therefore affects both the position and/or existence of the peak in question in addition to the integral values of all the other candidates (both through the altering of the popularity profile \textit{and} the integration procedure which maps it onto the discrete space of Fig.~\ref{fig:op0}). We will therefore need to find such hardness sets algorithmically. 

\vspace{-0.3cm}
\subsection{Metropolis Monte Carlo Search}
\begin{figure}[h]
\begin{center}
\resizebox{0.45\textwidth}{!}{\includegraphics[trim=1.5cm 1.2cm 0cm 1cm]{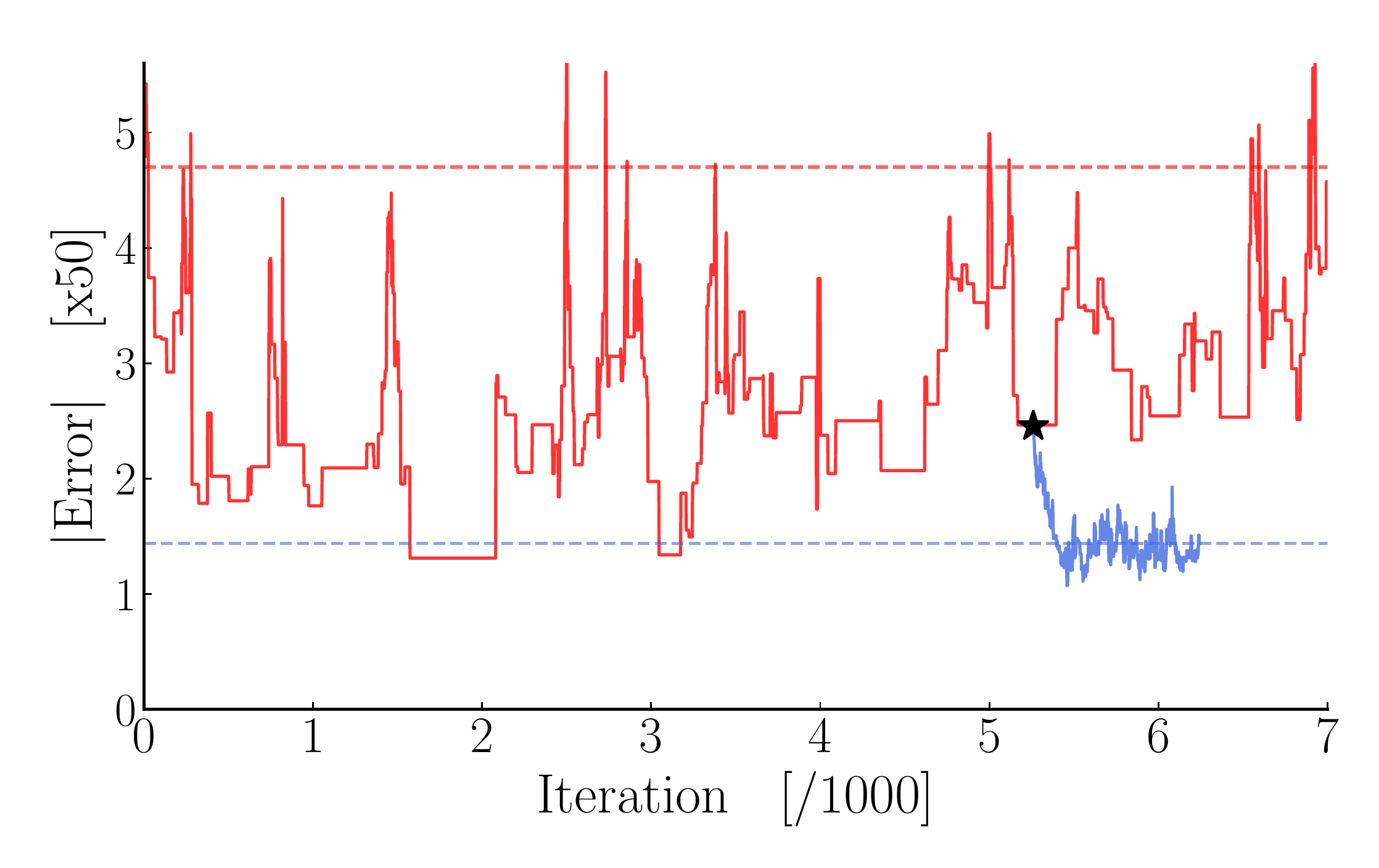}}
\end{center}
\caption{An example of a solution obtained by Monte Carlo. I will take the opportunity to christen this: Self-Consistent Heuristic Ordinal Opinion Loop Monte Carlo (SCHOOLMC). The red dashed line marks the error of the starting set. The red line is the `high temperature' search with the larger step size. Multiple basins can be observed. The blue line is just one of the many `low temperature' searches with the lower step size. The black star indicates where the algorithm has indentified a basin during the red search and therefore started the subsequent blue search. In this case the final error (blue dashed line) is higher than the lowest error found by the red line, and so this solution was rejected.} \label{fig:monty}
\end{figure}

\vspace{-0.1cm}
Without loss of generality ``May's Deal'' was fixed at -1. The domain was set to be 7 Brexit units either side of -1, under the assumption (backed up by tests) that sets which use larger ranges would not aid in producing meaningful orderings.\footnote{They tend to sacrifice a single vote in order to get good agreement with all the others. I think this is a manifestation of the fact that the model has an inherent floor to the error vector's size} The possible sets in the resulting seven-dimensional space were then explored using Metropolis Monte Carlo to minimize the size of the error vector. Firstly, a long run with large moves ([-1,1] for each candidate) and a low Metropolis multiplier (`high temperature') was performed. From this run low error sets were selected and a second run with a smaller range of allowed moves ([-0.1,0.1]) and a high Metropolis multiplier (4x as large, `low temperature') was performed for each. This is known as ``annealing''. An example of one set found in this way is shown in Fig.~\ref{fig:monty}. An error reduction of around 5$\times$ is the best that was observed.

Some of these sets are nonsensical, for example putting together ``Revoke Article 50'' and ``No Deal'' far off together at an extreme of the domain. Fortunately, many results constituted `sensible' sets, in which it was mainly the spacing from the original that had changed. By coding what it means to be `sensible' into the Monte Carlo in the form of some kind of Lagrangian, its efficiency could be massively increased, but I was wary of imposing too many of my own biases on the data at this stage. In any case, with no clear winner some amount of judgement must be employed. Certainly, more work could be done on the algorithm in general and these data in particular. 

\begin{figure}[h]
\begin{center}
\resizebox{0.5\textwidth}{!}{\includegraphics{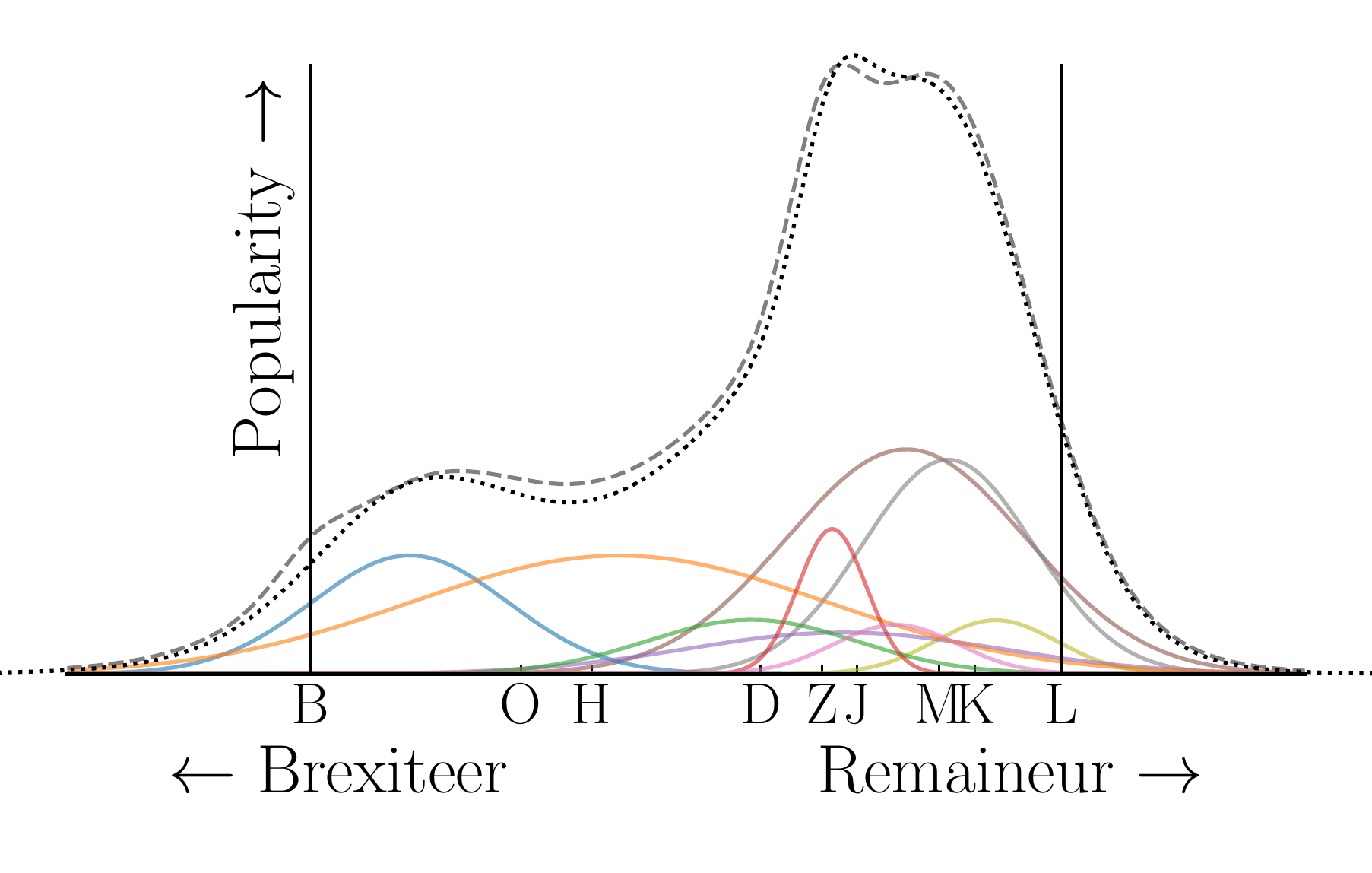}}
\resizebox{0.5\textwidth}{!}{\includegraphics[trim=1cm 1cm 1cm 1cm]{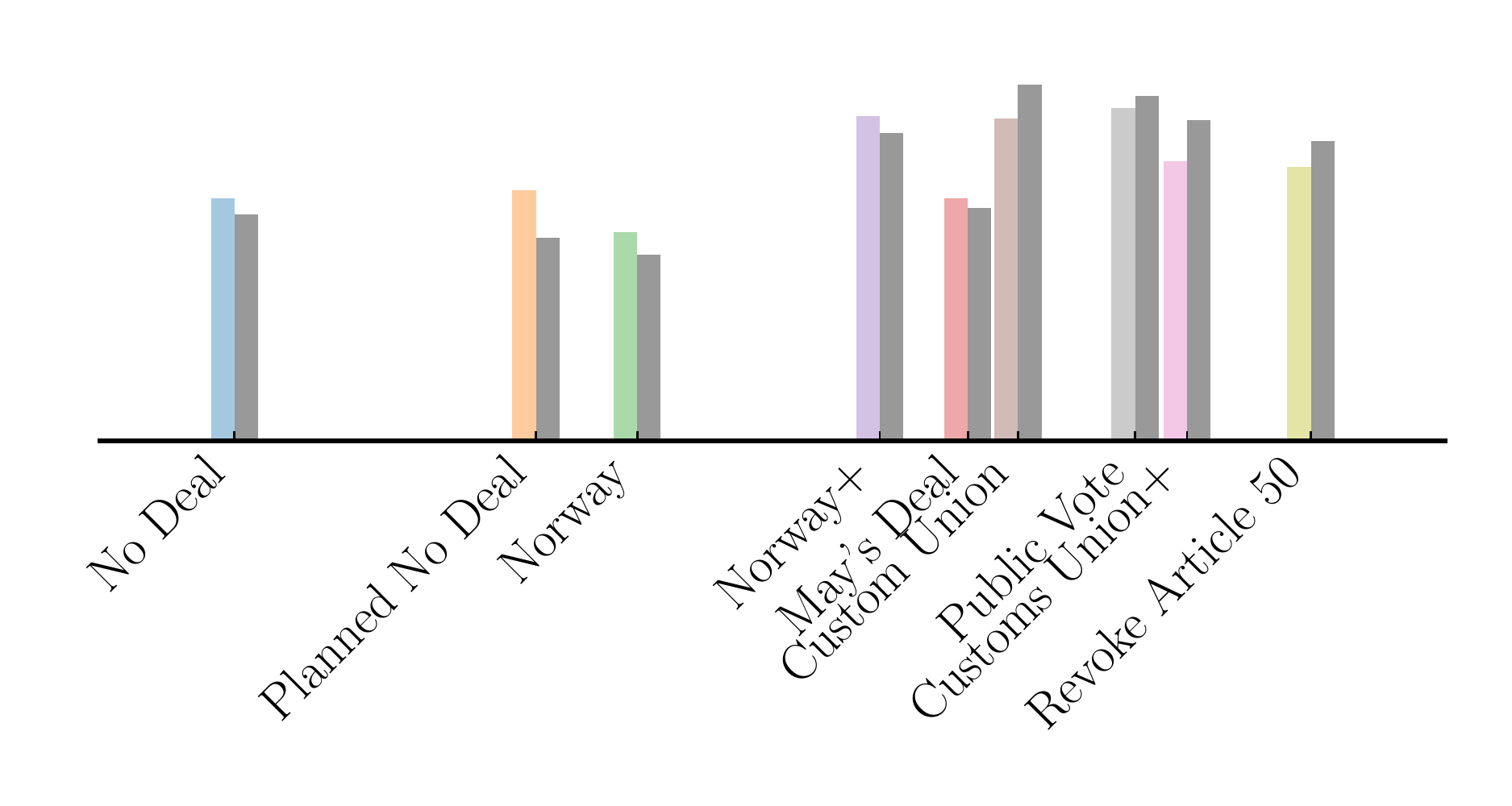}}
\end{center}
\caption{\textbf{Top}: Continuous popularity profile (as in Fig.~\ref{fig:op0fine}) for the optimised set. The x-axis has been rescaled, with the candidate's letter references (see Table~\ref{tb:cand}) displayed at their respective hardness values. ``May's Deal'' and ``Norway+'' have swapped in the ordering, as have ``Public Vote'' and ``Customs Union+''. \textbf{Bottom}: the corresponding discrete popularity profile (as in Fig.~\ref{fig:op0}, omitting dark-grey bars). The model claims ``Norway+'' is more popular than the original data suggest, and that ``Customs Union/+'' are less popular.} \label{fig:op1}
\end{figure}

For now we will concern ourselves with just one of these stochastically obtained choices for each candidate's hardness. The selected set's popularity profile is displayed in Fig.~\ref{fig:op1}, where the set itself is displayed as ticks below the axis. In this case, one interesting change is the ordering between ``Norway+'' and ``May's Deal''. We can understand this particular change as follows. Firstly, they both belong to the same peak of the total popularity distribution, and specifically are the two `most Brexit' options in that peak. ``Norway+'' is the more popular of the two. It follows that ``Norway+'' must be further to the edge of the cluster in order to gain the rump of the votes to the left, since votes to the right compete with the various `soft Brexit' options. The centre cannot hold. If this troubles the reader, perhaps because they see the broader appeal to Labour of ``Norway+'' as meaning that it must be `less Brexit' than ``May's Deal'', remember that this set is not a representation of where these candidate policies `actually are' on some `true' axis, but rather a suggestion of how they appeared to the parliamentarians when they were casting their votes. The course-grained version can be seen in Fig~\ref{fig:op1}, again with the observed data in grey for comparison, the large reduction in error from Fig.~\ref{fig:op0} is readily observable.

Several things stand out to the author. Firstly, the now two-peaked distribution and its large separation between ``No Deal'' and the other candidates is reminiscent of the Leave-Remain split of Fig.~\ref{fig:refer}. Parliament is much closer to the Remaineur extremists than the Leave extremists. Secondly, the set clusters options together in a way that is appealing given the discussion in Sec.~\ref{sec:rep}. Thirdly, ``May's Deal'' being located within the larger peak shows it is entirely consistent with the preferences of the majority-Labour groupings, as the Government often suggests.

\section{Hypothetical Ordinal Ballots}
One of the many curiosities of this particularly parliamentary episode was that its purpose was not particularly well-defined. The wording of the amendment which brought it about was,

\begin{itemize}
\item[] ``... and, given the need for the House to debate and vote on alternative ways forward, with a view to the Government putting forward a plan for the House to debate and vote on, orders that  Standing Order No. 14(1) (which provides that government business shall have
precedence at every sitting save as provided in that order) shall not apply on Wednesday
27 March...''\cite{Commons2019}
\end{itemize}
Most of the procedural fluff that followed concerned itself with obviating the government's attempts to thwart the exercise taking place at all.\cite{Business} In the event some candidate reached a majority, and that motion carried, it would not have had any particular legal force. Presumably it would also have been possible for multiple amendments to pass, which would have been even more interesting.

Although no candidate obtained a majority, the indicative votes still managed to produce a fascinating situation where, in the immediate aftermath, different politicians and pundits identified different candidates as the winner. Surely the winner was the candidate which obtained the most votes aye? Or perhaps it was the one which proved the most palatable, with the fewest votes no? Or instead was it the one which commanded the largest majority for aye over no?\cite{Alcantud2014} Delicious. Of course in reality, if push came to shove in a standard parliamentary vote, the large number of abstentees could (and probably would) trivially defeat any single candidate.

Only two things can end the Brexit saga. The first scenario is that somehow\footnote{And there are some exciting ways this could happen, from proroguing parliament tactically, to getting a member of the European Council to veto an extension} the legal default of ``No Deal'' occurs. The second is that a majority for one of the other options is forced. This might happen when the dark-horse of ``Revoke Article 50'' goes head-to-head against ``No Deal'' at the 11$^\rmm{th}$ hour. This concept of two options being run against each other directly was mentioned in the previous section. We shall now consider it explicitly by constructing the required candidate-orderings (known as \textit{ordinal ballots}) from our logical MPs. That is to say we will use the results of the cardinal voting to predict result of an ordinal vote.

\subsection{The Condorcet Process}
For a particular hardness set, it is possible to simulate the result of a 1-on-1 contest between two options by assigning each MP to the candidate who is closest to their position on the Brexit axis. We will not allow abstentions, though they can clearly affect the result (this is the \textit{No Show Paradox},\cite{Moulin1988} more on this in a moment). The outcomes of our electoral equivalent of trial by combat for our two hardness sets are shown in Table~\ref{tb:cond1}. In the original set, ``Norway+'' emerges as the Condorcet winner. In the optimised set, the victor is ``Customs Union''. Thankfully neither produce a tie. This is only a simulation, but what it does show is that positions directly adjacent to ``May's Deal'' on the Brexit axis are not only more popular, but are popular enough to actually be able to defeat every single other candidate. The tables are also useful in that they predict the result of any particular head-to-head one wished to focus on, as in the `Revoke Article 50'' vs. ``No Deal'' example: ``Revoke Article 50'' won narrowly for the original set, and more convincingly with the optimised set.

\begin{table}[h]
\begin{tabular}{|c| |c|c|c|c|c|c|c|c|c| |c| }
\hline
\textbackslash & B & O & H & Z & D & J & K & M & L & Wins\\\hline\hline
B & 0 &   60 &   71&  141&  186&  215&  268&  290&  304 & 0 \\\hline
O & 577 &    0 &   98&  173&  190&  232&  280&  299&  323 & 2 \\\hline
H & 566 &  539 &    0&  190&  232&  280&  299&  323&  359 & 4 \\\hline
Z & 496 &  464&  447&    0&  280&  299&  323&  359&  443 & 6 \\\hline
\textcolor{Purple}{D} & \textcolor{Purple}{451} &  \textcolor{Purple}{447}&  \textcolor{Purple}{405}&  \textcolor{Purple}{357}&    \textcolor{Purple}{0}&  \textcolor{Purple}{323}&  \textcolor{Purple}{359}&  \textcolor{Purple}{443}&  \textcolor{Purple}{538} & \textcolor{Purple}{8} \\\hline
J & 422 &  405&  357&  338&  314&    0&  443&  538&  603 & 7 \\\hline
K & 369 &  357&  338&  314&  278&  194&    0&  603&  620 & 5 \\\hline
M & 347 &  338&  314&  278&  194&   99&   34&    0&  623 & 3 \\\hline
L & 333 &  314&  278&  194&   99&   34&   17&   14&    0 & 1 \\\hline
\end{tabular}

\vspace{0.5cm}

\begin{tabular}{|c| |c|c|c|c|c|c|c|c|c| |c|}
\hline
\textbackslash & B & O & H & D & Z & J & M & K & L & Wins\\\hline\hline
B& 0&   12&   67&   70&  120&  142&  172&  178&  192 & 0\\\hline
O& 625&    0&   97&  178&  186&  192&  219&  223&  248 & 1\\\hline
H& 570&  540&    0&  190&  193&  212&  244&  246&  286 & 2\\\hline
D& 567&  459&  447&    0&  248&  254&  307&  319&  432 & 5\\\hline
Z& 517&  451&  444&  389&    0&  291&  323&  388&  527 & 7\\\hline
\textcolor{Bittersweet}{J}& \textcolor{Bittersweet}{495}&  \textcolor{Bittersweet}{445}&  \textcolor{Bittersweet}{425}&  \textcolor{Bittersweet}{383}&  \textcolor{Bittersweet}{346}&    \textcolor{Bittersweet}{0}&  \textcolor{Bittersweet}{387}&  \textcolor{Bittersweet}{447}&  \textcolor{Bittersweet}{582} & \textcolor{Bittersweet}{8}\\\hline
M& 465&  418&  393&  330&  314&  250&    0&  563&  623 & 6\\\hline
K& 459&  414&  391&  318&  249&  190&   74&    0&  637 & 4\\\hline
L& 445&  389&  351&  205&  110&   55&   14&    0&    0 &3\\\hline
\end{tabular}

\vspace{0.2cm}

\caption{This is to be read as, `Row [letter1] won $\rightarrow$ votes against $\uparrow$ column [letter2]. \textbf{Top}: Condorcet election simulated using the MPs' mean score under the original hardness set to construct their ordinal ballots. D$\equiv$``Norway+'' is the Condorcet winner. \textbf{Bottom}: Condorcet election simulated using the MPs' mean score under the optimised hardness set to construct their ordinal ballots. J$\equiv$``Customs Union'' is the Condorcet winner.} \label{tb:cond1}
\end{table}

It would be beyond exciting to see this sort of vote implemented in the House of Commons. However a word of warning. Our logical MPs both vote sincerely and fill out their entire ballot. One could instead vote tactically, either by abstaining to vote or, more likely, attempt to defeat other candidates that one considered contenders. This is known as \textit{burying}. For example, in Table~\ref{tb:cond1}, top, D beats J overall. Large numbers of J voters could vote for Z ahead of D even though this contrary to their sincere preference. If Z beats D then J ties for first place with Z, instead of coming second! One could imagine tactical voting being especially effective (and exciting) in such a small electorate where the preferences of the other electors can be fairly accurately gauged. Game theory will then ensue. The resistance to this sort of tactical voting is the subject of some amount of literature which I am not able to penetrate with any success at this point. Nevertheless, as far as I can tell, this sort of vulnerability is brought about by the Condorcet process not fulfilling the \textit{later-no-harm criterion}. That is to say exactly the D--J--Z situation described above. However, there exist other voting systems which do not exhibit this particular flaw.

\subsection{Instant Runoff}

If you are interested in a poorly thought-out referendum on a complex issue with corrupt financing, misleading claims and/or promises, and an ultimate motivation to maintain a Conservative majority, then what better topic than Alternative Vote! If there was one issue that I do not recall being extensively debated in that referendum campaign, it was that Alternative Vote (AV) does not necessarily select the Condorcet winner, as described above. However, it does fulfil the later-no-harm criterion. More on that in just a moment.

\begin{figure}[h]
\begin{center}
\resizebox{0.5\textwidth}{!}{\includegraphics[trim=0cm 2.13cm 0cm 0.7cm]{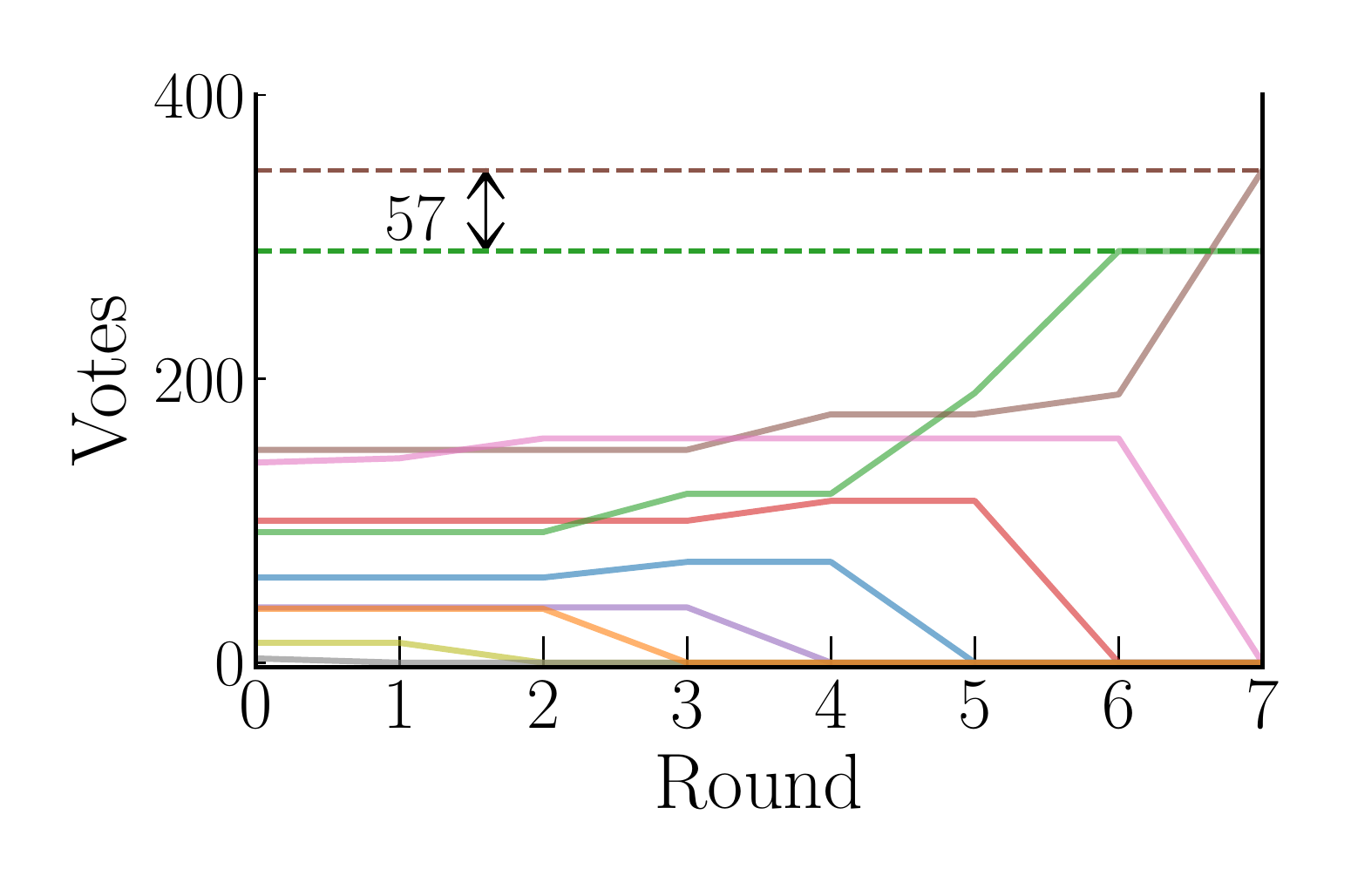}}
\resizebox{0.5\textwidth}{!}{\includegraphics[trim=0cm 2cm 0cm 0.5cm]{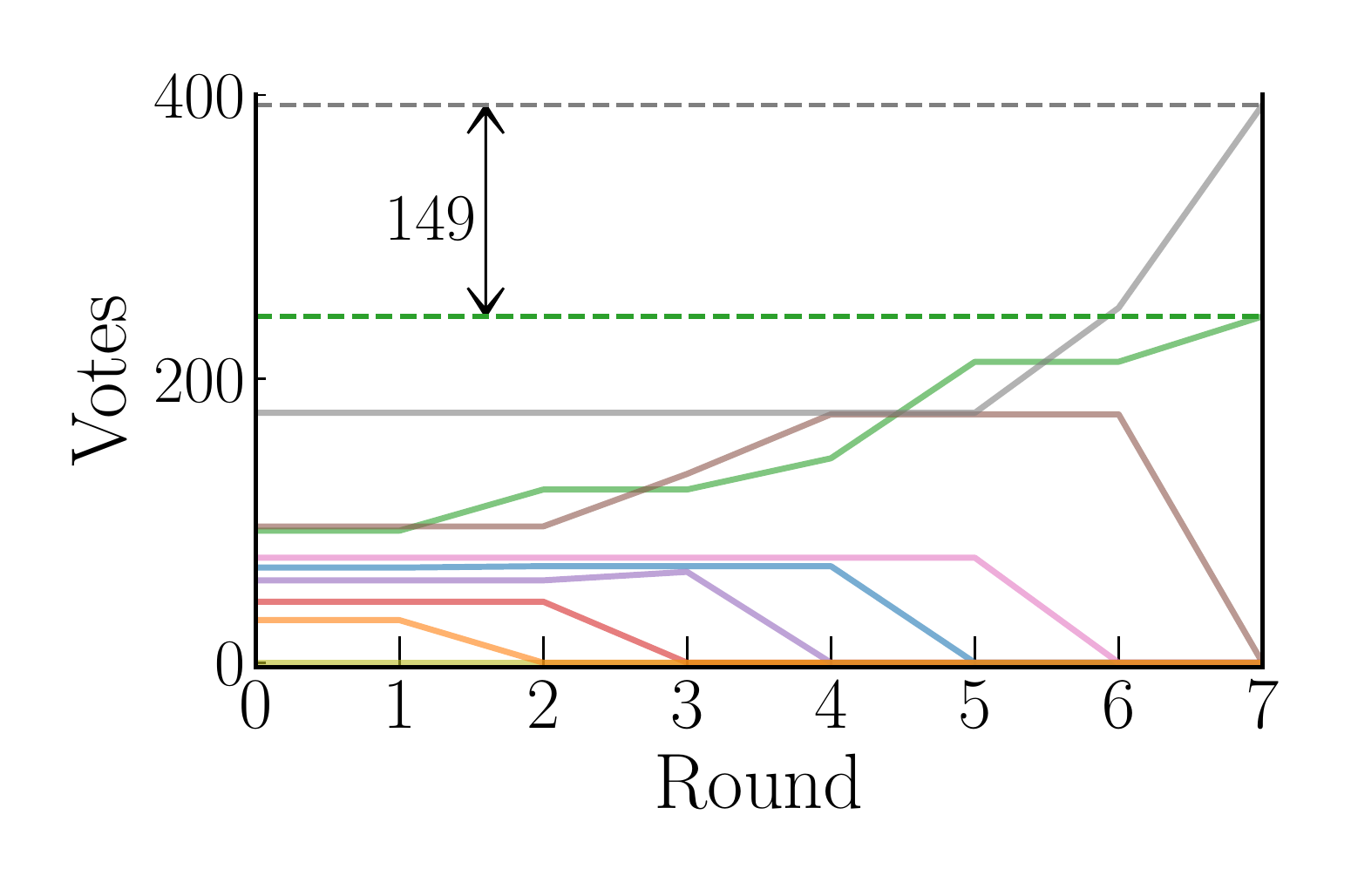}}
\end{center}
\caption{\textbf{Top}: Alternative Vote simulated using the MPs' mean score under the original hardness set to construct their ordinal ballots. ``Customs Union'' beats the Brexiteer candidate of ``Norway'' after ``Customs Union+'' is knocked out in the penultimate round. ``May's Deal'' comes fourth. \textbf{Bottom}: Alternative Vote simulated using the MPs' mean score under the optimised hardness set to construct their ordinal ballots. ``Public Vote'' trounces ``Norway'' after the defeat of ``Customs Union''. The three had been neck-and-neck for some time. ``May's Deal'' comes 7$^\rmm{th}$.} \label{fig:av1}
\vspace{-0.5cm}
\end{figure}

For those who do not remember, or weren't paying attention at the time, in AV one ranks the candidates in order of preference. The ballots are then counted and the candidate with the fewest top-choice votes is eliminated. The count is then performed iteratively, with the eliminated candidate deleted from everyone's ballots. Such an order of preferences can be constructed from these data in the same way as for the Condorcet process, we can then simulate the result of the candidates contesting an AV election. This process always returns a winner. If it could be agreed upon, it would therefore be able to bring about a definitive end to the Brexit debate. In the original set, we found that ``Customs Union'' defeated ``Norway'' in the final round with a majority of 57; for the optimised set, ``Public Vote'' defeated  ``Norway'' in the final round with a much larger majority of 149, see Fig.~\ref{fig:av1}.

Can this process also be manipulated by tactical voting? The answer is yes, to this and in fact to \textit{all} deterministic voting systems you could possibly conceive by something known as the Gibbard--Satterthwaite theorem! A description is particularly relevant if ``Public Vote'' were to win. If a second referendum were to be conducted, everyone has a different idea about what would be on the ballot. Certainly, if the ballot read ``No Deal'', ``May's Deal'', ``Revoke Article 50'', it could be considered rather unfair! Since the Leave vote, which had only a small majority, would be divided between the first two and ``Revoke Article 50'' would cruise to victory. This is a generic weakness of plurality voting which leads to the sort of tactical voting we see in UK general elections. It is logical, therefore, for a referendum between these 3 options to be subject to runoff. This would not `split' the Leave vote in the sense that it would correctly be able to represent the order of preferences of each individual. However, it suffers from a violation of what is known as the \textit{monotonicity criterion}. This is in a sense a more extreme example of the later-no-harm criterion, because it requires you to not vote for your preference first. This is best explained with an example.

\begin{itemize}
\item[1)] Firstly we consider a sincere vote. In the first round ``Revoke Article 50'' comes top, as it would in a single-round plurality voting process. ``May's Deal'' comes last, and the votes are redistributed. Not enough of the ``May's Deal'' vote transfers to ``No Deal'', so ``Revoke Article 50'' still wins. This would not be the Condorcet winner, since in a straight fight ``Revoke Article 50'' would lose to ``May's Deal''. This can be exploited. 
\item[2)] Now consider a tactical vote. ``No Deal'' conspirators have foreseen the outcome in the sincere scenario, of course. Determined to stop Remain at any cost, they raise ``May's Deal'' to the top of their ballots. ``Revoke Article 50'' still wins in the first round. However, when ``No Deal'' now loses instead as result of the conspirators' actions, their votes redistribute to ``May's Deal''. The head-to-head results in ``May's Deal'' winning. 
\end{itemize}

\pagebreak
This strategy is known as \textit{compromise}. It was never possible for the ``No Deal'' conspirators to win, and so they spited their nemesis by voting for some 3$^\rmm{rd}$ option. This seems like a perfectly reasonable thing to do, since the benefactor of your tactic is not yourself. It is effectively what occurred in the 3$^\rmm{rd}$ vote on ``May's Deal'' in the Commons. However, for other groups, or in other electorates, the strategy can actually elect your own candidate at the end. This is then termed \textit{pushover}. A slightly convoluted example of this in the current electorate:

\begin{itemize}
\item[3)] Return to the second example with the compromise victory. The ``Revoke Article 50'' camp predict this eventuality and seek to thwart it. They know that ``No Deal'' will lose to them, i.e. it is a weak candidate, a `pushover'. Thus, in the first round they raise ``No Deal'' to the top of their ballots, cancelling out the defection of the ``No Deal'' conspirators to ``May's Deal'' and ensuring ``No Deal'' still comes second. The final result is then the same as in the first case, and ``Revoke Article 50'' wins in spite of people expressly reducing the position of it on their ballots. ``No Deal'' should be an \textit{irrelevant alternative}, but its position in the ballot turns out to influence the result between the two `relevant' competitors.

If that's a bit too much of a train-wreck of gambits for the reader to entertain, one can instead consider that the ``Revoke Article 50'' camp merely believe that ``No Deal'' is set to come last. The mechanics are the same, except, of course, that as a result of their very attempt to manipulate the election to their advantage, ``No Deal'' could be elected accidentally! This element of risk is a defence against vote-manipulation based on the non-monotonic behaviour of AV, as the process threatens to descend into a glorified game of rock-paper-scissors.
\end{itemize}

\begin{figure}[h]
\begin{center}
\resizebox{0.5\textwidth}{!}{\includegraphics[trim=0cm 2.1cm 0cm 0.5cm]{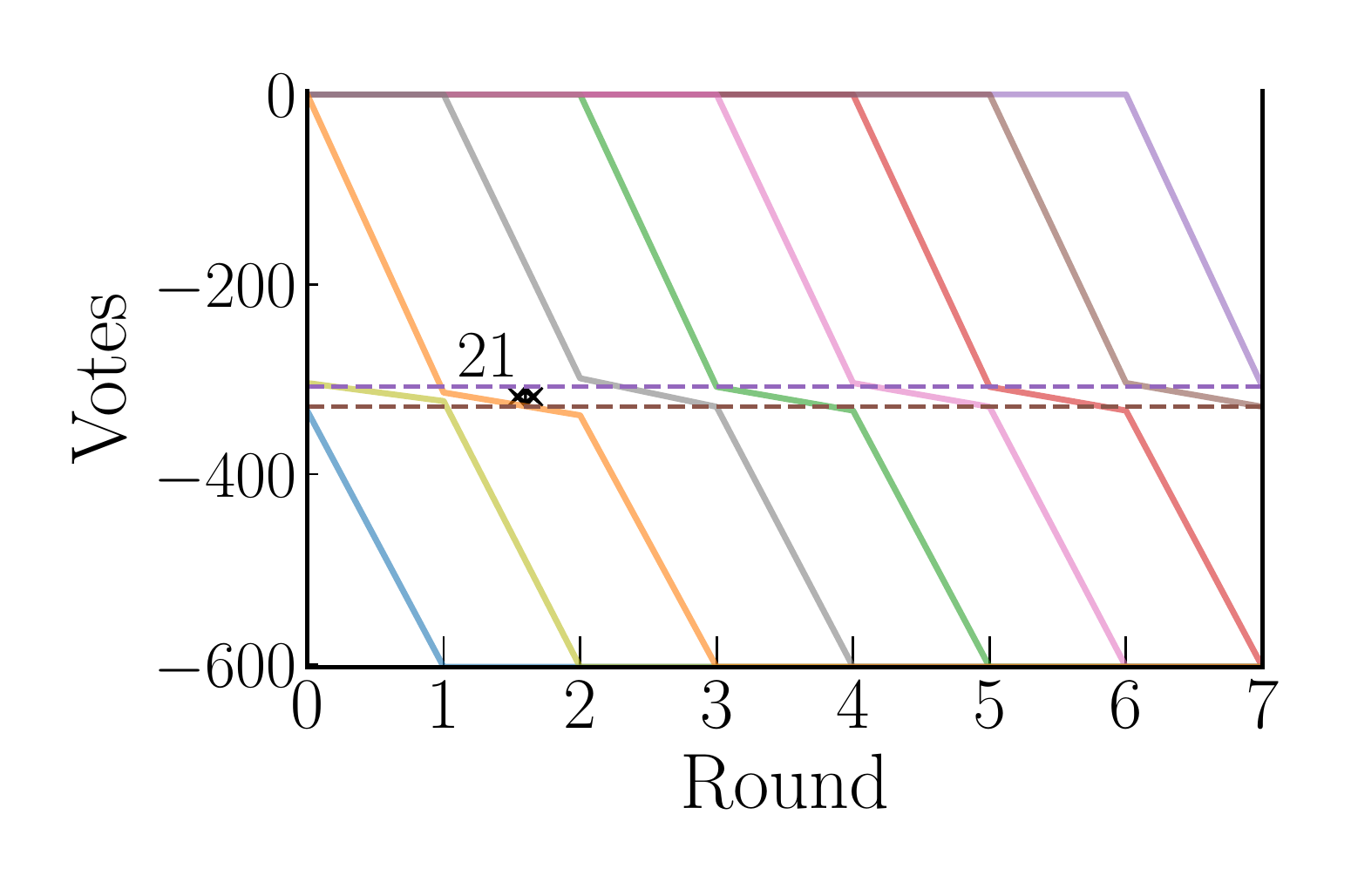}} \label{fig:coombs0}
\resizebox{0.5\textwidth}{!}{\includegraphics[trim=0cm 2cm 0cm 0.5cm]{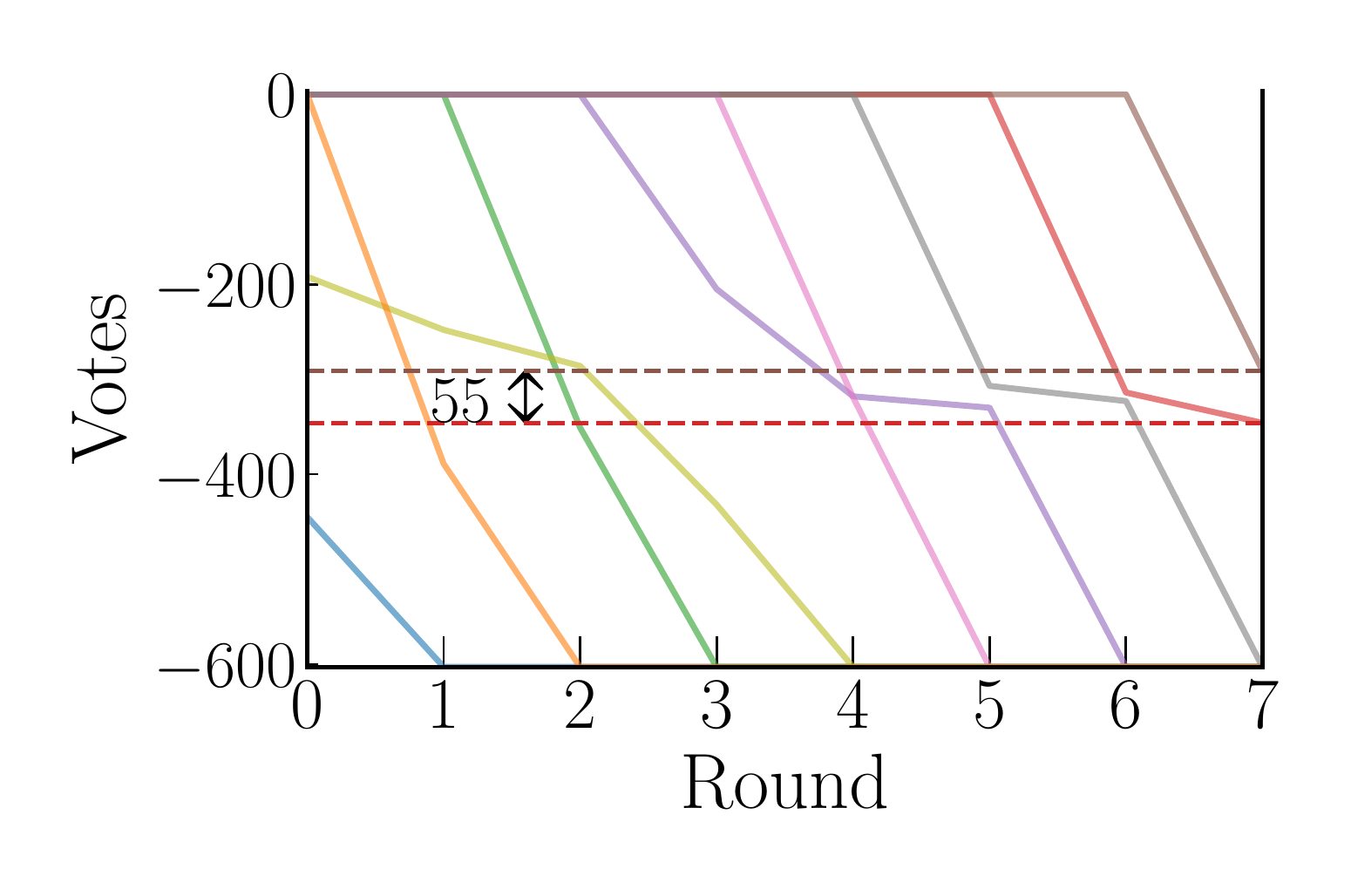}}
\end{center}
\caption{The number of last-place nominations is displayed at each stage as a number of negative votes. After being eliminated, candidates are displayed with -600 votes. \textbf{Top}: Coombs Rule simulated using the MPs' mean score under the original hardness set to construct their ordinal ballots. ``Norway+'' narrowly defeats ``Customs Union''. ``May's Deal'' comes third. \textbf{Bottom}: Coombs Rule simulated using the MPs' mean score under the optimised hardness set to construct their ordinal ballots. ``Customs Union'' defeats ``May's Deal'' after it narrowly avoids elimination to ``Public Vote''. The behaviour is considerably more interesting.} \label{fig:coombs1}
\vspace{-0.6cm}
\end{figure}

While we are considering hypothetical runoff ballots, it is worth exploring something called Coombs' Rule. This is back-to-front AV in which the most least-popular candidate is eliminated in each round, instead of the least most-popular. The winner is therefore the candidate who the least people do not want, as opposed to the one most people actively want. Recall that it was mentioned that some pundits claimed their preferred option had won on those grounds? Well this is a method to find the true candidate which fulfils that property. The process shares many of the same properties as AV, an analysis of which is available.\cite{Grofman2004} In the original set, ``Norway+'' defeats ``Customs Union'' in the final round with a majority of 21; in the optimised set, ``Customs Union'' defeats  ``May's Deal'' in the final round with a majority of 55, see Fig.~\ref{fig:coombs1}. It is worth noting that in this case, the Coombs' Rule produces the Condorcet winner. Indeed, one can imagine that burying would be a good tactical strategy on this sort of ballot. In these two areas it is more similar to Condorcet than to AV.

\vspace{-0.1cm}
\section{Conclusion}
\vspace{-0.1cm}

In this paper we have constructed a 2D representation of MPs' attitudes. From these data we have conceived of a self consistent procedure that used Monte Carlo sampling to define an `objective' hardness set for the candidate policies. One such set was considered in the analysis that followed. This set appealed due to its: clustering of options perceived as similar, expression of the polarized nature of the debate, the fact that Labour's behaviour is to some degree motived by partisan tendencies rather than preferences. In the final section we hypothesised on what the results of three kinds of ordinal ballot would have been, if they had been carried out. Both the original set and the optimised set were compared. Interestingly, a pure Condorcet process returned a winner in both cases. This is probably a result of the logical MPs being unable to express cyclical preferences, though (and this is really weird) that does not ensure the aggregate preferences are not cyclical.\cite{WilliamV2002} In the case of the optimised set, the Coombs Rule produced the Condorcet winner.

\begin{figure*}[t]
\begin{center}
\resizebox{0.8\textwidth}{!}{\includegraphics[trim=0cm 1cm 0cm 1.5cm]{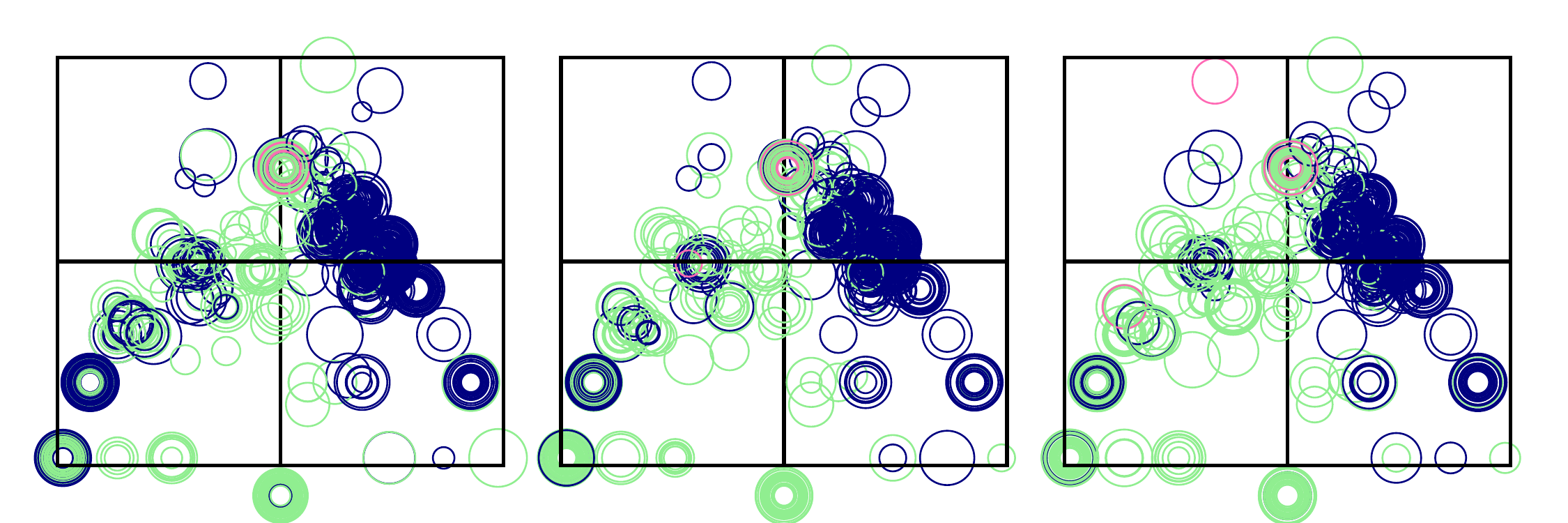}}
\end{center}
\caption{Fig.~\ref{fig:original} where MPs are coloured according to how they voted in the three votes on ``May's Deal'', ordered chronologically from right to left. Light blue: aye, dark blue: no, pink: no vote recorded. We used data from the first two where there was a perceived threat of ``No Deal'' in the paper.} \label{fig:mayvotes}
\end{figure*}

\pagebreak
How does this help inform the debate? Ultimately, Mrs. May's chosen deal will receive more votes than predicted in Sec.~\ref{sec:cardinal}, as it did on her third attempt (the day after the indicative votes, Fig.~\ref{fig:mayvotes}). This is partly owing to her use of the whip, and partly her threat to parliament that if the vote should not pass, the alternative might be worse. In some sense, this is equivalent to running ``May's Deal'' against whatever the threat was (``No Deal'' in the first two cases, and perhaps ``Revoke Article 50'' in the third case) in a head-to-head. Of course, the effectiveness then depends on the credibility of that threat, which we could perhaps address using the logical MPs. This could constitute an interesting extension. In any case, the thrust of the argument is that the government's chosen position is privileged. Since ``Norway+'' and ``Customs Union'' (the neighbouring candidates for the two sets we considered, respectively) were found to be both the Condorcet and Coombs winners, it follows that the Prime Minister need only move a small amount from her current position in order to carry a large majority. This will, of course, involve crossing at least one of her `red lines'. What these results show with absolute conviction, however, is that they must indeed be crossed. Only policies which do this are able to result in victory under this model. None of the conceived ballots show ``May's Deal'' coming out on top, and all of the options to the left of her (harder Brexits) do even worse. This is ultimately a manifestation of the `Remain parliament'.

On an authorial note: I am unable to tell to what extent the work presented in this paper represents a methodological novelty. If it does, then the next step would be to test it on data sets where both cardinal and ordinal ballots were conducted in parallel. If it does not, then the author would be enormously grateful if he could be made aware of the relevant papers.

\begin{acknowledgments}
TS is supported by a departmental studentship (No.~RG84040) sponsored by the Engineering and Sciences Research Council (EPSRC) of the United Kingdom, who probably won't be happy he spent a week on this. Conor Heffernan and Hugh Burton are thanked for helpful discussions. Jacqueline O'Hara is thanked for reading the manuscript. The reader is thanked for putting up with my unmollified literary style long enough to reach the acknowledgements.
\end{acknowledgments}
\vspace{4cm}

\section{References and Footnotes}
\bibliography{Brexit}

\appendix
\section{Additional Figures}
See overleaf for those figures mentioned in, but not essential to, the main body of text.
\vfill
\begin{figure*}[b]
\begin{center}
\begin{minipage}{0.45\textwidth}
\resizebox{0.7\textwidth}{!}{\includegraphics[trim=0cm 0cm 0cm 0.5cm]{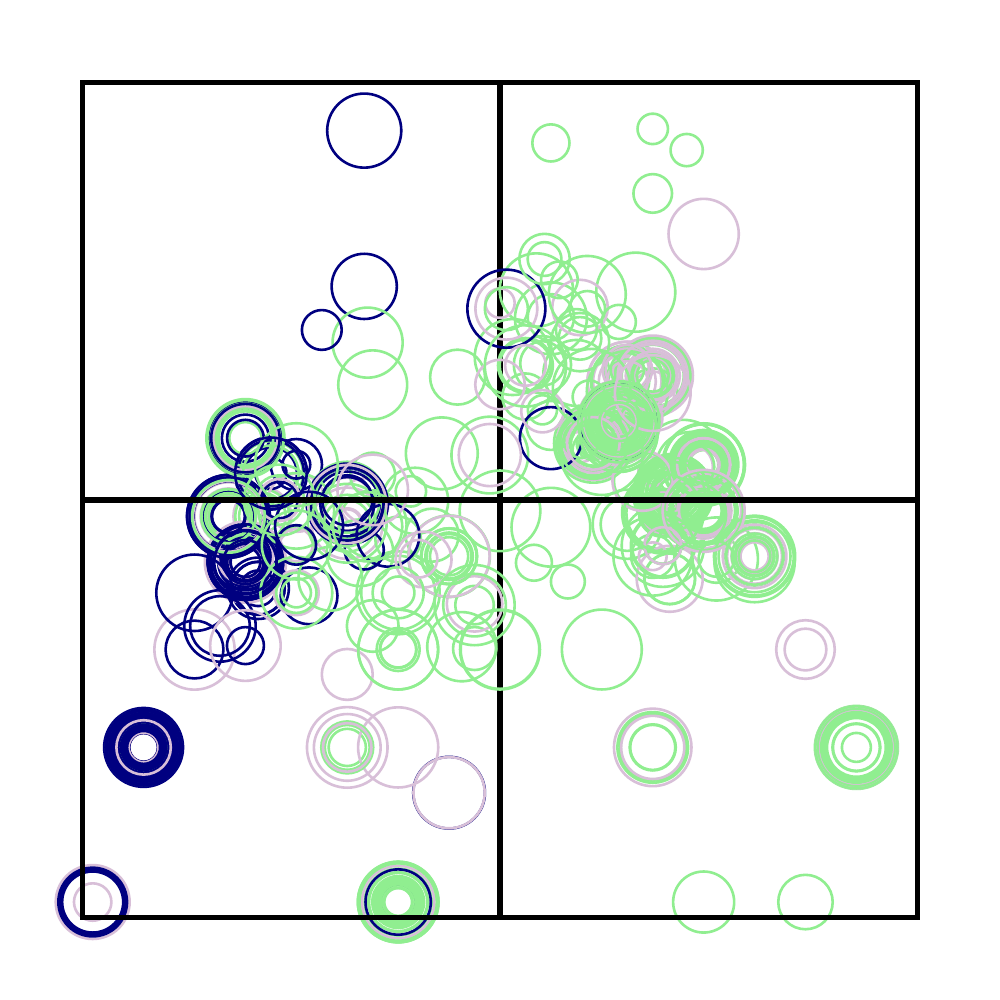}}
\caption{Fig.~\ref{fig:bigplot} colour coded according to stance taken in the 2016 referendum. Dark blue: Leave, light blue: Remain, lilac: was not an MP at the time or did not publically declare. The data show firstly that parliament has a large majority of Remain supporters, and secondly that the most Leave supporters cluster around the left-most extreme.} \label{fig:refer}
\end{minipage}\hfill
\begin{minipage}{0.45\textwidth}
\resizebox{0.6\textwidth}{!}{\includegraphics[trim=2cm 1cm 2cm 0.5cm]{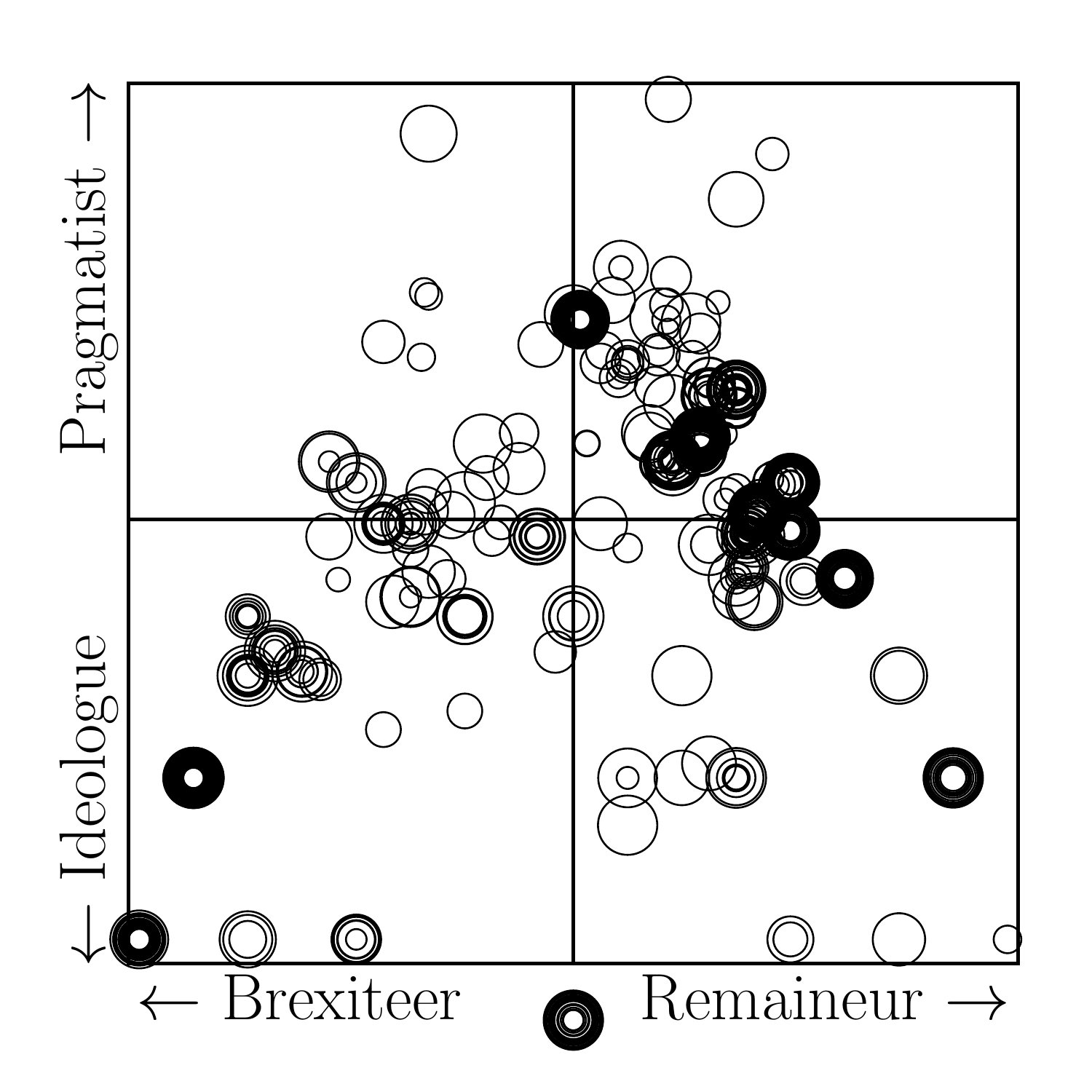}}
\caption{Original data without the inclusion of the votes on ``May's Deal''. Those who voted no to everything are below the axes. Huw Merriman is still not being displayed for clarity (he has an even larger variance than in the text). The cluster just to the right of center with a high variance are those who abstained on all options: the Cabinet, the Speaker's office, Sinn F\'{e}in, and one SNP MP.} \label{fig:original}
\end{minipage}

\end{center}
\end{figure*} 

\end{document}